\documentstyle[12pt,epsf]{article}
\newcommand {\beq}{\begin{equation}}
\newcommand {\eeq}{\end{equation}}
\newcommand {\beqa}{\begin{eqnarray}}
\newcommand {\eeqa}{\end{eqnarray}}
\newcommand {\n}{\nonumber \\}

\newcommand {\Ione}{\mbox{\scriptsize I}}
\newcommand {\IItwo}{\mbox{\scriptsize II}}
\newcommand {\IItwoa}{\mbox{\scriptsize II(a)}}
\newcommand {\IItwob}{\mbox{\scriptsize II(b)}}
\newcommand {\wb}{\mbox{\scriptsize WB}}
\newcommand {\ee}{\mbox{e}}

\newcommand {\dd}{\mbox{d}}

\newcommand {\del}{\partial}

\newcommand {\defeq}{\stackrel{\rm def}{=}}

\begin{document}
\setlength{\oddsidemargin}{0cm}
\setlength{\baselineskip}{7mm}

\begin{titlepage}
 \renewcommand{\thefootnote}{\fnsymbol{footnote}}
%\vspace{-3cm}
\begin{normalsize}
\begin{flushright}
\begin{tabular}{l}
UTHEP-399\\
NBI-HE-99-02\\
%hep-th/???????\\
February 1999
\end{tabular}
\end{flushright}
  \end{normalsize}

%~~\\
\vspace{1cm}

\vspace*{0cm}
    \begin{Large}
       \begin{center}
{Translational Anomaly in Chiral Gauge Theories}\\
{on a Torus and the Overlap Formalism}
%{Absence of an Anomaly Free U(1) Chiral Gauge Theory}\\
%{on a Two-Dimensional Torus}
%{Absence of an Anomaly Free U(1) Chiral Gauge Theory}\\
%{on a Two-Dimensional Torus}
%         {An ambiguity in the Wigner-Brillouin phase choice}\\
%{in the overlap formalism for lattice chiral gauge theory} \\
% the What the Overlap Formalism Can Do and What It Fails to Do}      \\
       \end{center}
    \end{Large}
\vspace{1cm}

%~~\\

\begin{center}
           Taku I{\scshape zubuchi}$^{1)}$\footnote
            {
e-mail address : 
izubuchi@het.ph.tsukuba.ac.jp}
           {\scshape and}
           Jun N{\scshape ishimura}$^{2)}$\footnote
           {
e-mail address : nisimura@alf.nbi.dk}\\
      \vspace{1cm}
        $^{1)}$ {\itshape Institute of Physics, University of Tsukuba,}\\
                 {\itshape Ten-oh-dai,Tsukuba 305-8571, Japan}\\
        $^{2)}$ {\itshape Department of Physics, Nagoya University,}\\
               {\itshape Chikusa-ku, Nagoya 464-8602, Japan} \\
        $^{2)}$ {\itshape Niels Bohr Institute, Copenhagen University,} \\
              {\itshape Blegdamsvej 17, DK-2100, Copenhagen \O, Denmark}
%      \vspace{1cm}
\end{center}

\hspace{5cm}

%\vfill

\begin{abstract}
\noindent
We point out that 
a fermion determinant of a chiral gauge theory
on a 2D torus has a phase ambiguity 
proportional to the Polyakov loops
along the boundaries,
which can be reproduced by the overlap formalism.
We show that the requirement on the fermion determinant
that a singularity in the gauge field
can be absorbed by a change of the boundary condition for the fermions,
is not compatible with translational invariance in general.
As a consequence,
the gauge anomaly for singular gauge transformations
discovered by Narayanan-Neuberger actually exists in any 2D U(1) chiral
gauge theory unless the theory is vector-like.
We argue that the gauge anomaly is peculiar to the overlap formalism
with the Wigner-Brillouin phase choice and that it is not necessarily
a property required in the continuum.
We also generalize our results to any even dimension.
\end{abstract}
\vfill
\end{titlepage}
\vfil\eject

\setcounter{footnote}{0}

\section{Introduction}
\setcounter{equation}{0}
\renewcommand{\thefootnote}{\arabic{footnote}}

The overlap formalism \cite{NN}
is one of the most promising approaches
to lattice chiral gauge theories.
There are a number of tests that have been done so far.
%and all of them are reported to be successful.
For fixed gauge backgrounds,
the perturbative anomaly \cite{PertAnom}
and the vacuum polarization \cite{VP}
have been reproduced analytically.
Also exact chiral determinants for a 2D 
U(1) chiral gauge theory 
have been reproduced \cite{torus,FRD,Fosco,Anomaly}
for antiperiodic boundary conditions.
Some attempts have been made to test the formalism
with the dynamical gauge field, and
%it is reported that
the analytic result for the 't Hooft vertex 
has been correctly reproduced \cite{Composite,Finite,MCeval}
by simply averaging over the gauge orbit in the path integral
of the gauge field.

The formalism is not restricted to chiral gauge theories,
but it can be applied to 
any kind of fermion on the lattice, where
exact symmetries of the formalism
are of advantage over conventional approaches
to lattice fermions.
When applied to Dirac fermions in even dimensions, 
an exact chiral symmetry is preserved and
one can even derive a Dirac operator \cite{exactly},
which gives an explicit solution \cite{more} 
to the Ginsparg-Wilson relation \cite{GW}.
When applied to Dirac fermions in odd dimensions,
parity invariance can be manifestly preserved
\cite{NarayananNishimura},
and a global gauge anomaly can be correctly reproduced 
\cite{KikukawaNeuberger}.
These symmetries enable a lattice construction of
supersymmetric gauge theories without fine-tuning \cite{NN,MN}.
Realizing lattice supersymmetry for the free case 
is also succeeded \cite{AK}.

In this paper, 
we investigate the overlap formalism
as a lattice construction of chiral gauge theories.
We point out that in chiral gauge theories on a two-dimensional 
torus,
the fermion determinant has a phase ambiguity proportional to 
Polyakov loops along the boundaries, 
which does not exist in vector-like gauge theories.
This generally gives rise to a translational anomaly in these theories.
The ambiguity can be 
reproduced by the overlap formalism
as an ambiguity in the choice of 
the boundary condition for the reference state
used in the Wigner-Brillouin phase choice.
Imposing the translational invariance
corresponds to taking 
the boundary condition to be identical to the one for the 
fermion under consideration.
In this case, however,
a singularity in the gauge field
cannot be absorbed by a change of the boundary condition for the fermions
without having an extra phase factor to the fermion determinant.

This fact leads to the conclusion that
no matter how one fixes the phase ambiguity of the fermion determinant,
the anomaly for singular gauge transformations discovered in
Ref. \cite{Anomaly} cannot be cancelled unless the theory is vector-like.
As pointed out in Ref. \cite{Anomaly},
2D U(1) chiral gauge theories can have an anomaly 
for singular gauge transformations, 
even if the fermion contents and the boundary
conditions are chosen such that the gauge anomaly 
for non-singular gauge transformations is already cancelled.
The phase choice of the fermion determinant
with which the issue was discussed in
Ref. \cite{Anomaly} actually 
corresponds to taking the boundary condition for the reference state
to be antiperiodic irrespective of the boundary conditions
for the fermions under considerations.
%, which breaks the translational invariance.
In order to cancel the anomaly under singular gauge transformations,
%and thus making a completely anomaly-free chiral gauge theory,
it was proposed to take a special boundary condition for each fermion species.
The argument was restricted to the case when
the singularity lies exactly on the boundary, where the boundary
condition is imposed.
However, the translational invariance is broken with that phase choice,
which means that we also have to consider a more generic case
in which the singularity lies off the boundary.
We then find that the anomaly for singular gauge transformations
cannot be cancelled unless the theory is vector-like.
%One would like to reconsider this issue with the most natural phase choice
%with the translational invariance.
The conclusion actually does not depend on 
how one fixes the phase ambiguity,
and in particular, it remains unchanged 
for the translationally invariant phase choice.

We also generalize our results 
%concerning the translational anomaly and the singular gauge anomaly
to any even dimension in the abelian case.
At first sight, 
the existence of the singular gauge anomaly seems 
to contradict the fact that there exists
an explicit construction
of lattice U(1) chiral gauge theory,
which is gauge invariant on the lattice \cite{Luescher}.
This apparent contradiction can be solved by noting that
there is actually an ambiguity in the continuum calculations
of the chiral determinants for singular gauge configurations.
%This means that the continuum calculations alone cannot
%exclude the possibility that a completely gauge invariant 
%lattice regularization exists for chiral gauge theories.
The anomaly under singular gauge transformations
is a property of the overlap formalism with the Wigner-Brillouin
phase choice, but it is not
necessarily a property required in the continuum.

This paper is organized as follows.
In Section \ref{Review},
we briefly review 2D U(1) chiral gauge theory,
which we use for any explicit calculation of the 
fermion determinants.
The subtlety which gives rise to translational anomaly is revealed.
%We point out that translational invariance can be broken
%if one makes a change of the boundary condition for fermions
%by introducing a singularity in the background gauge field.
In Section \ref{Overlap}, we review the overlap formalism
and explain the ambiguity of the formalism.
%, which we discuss in this paper.
In Section \ref{BC}, 
we examine the continuum limit of the overlap formalism
for the 2D U(1) case.
We show that the ambiguity of the formalism gives
a phase ambiguity of the fermion determinant
proportional to the Polyakov loops along the boundaries.
%and that the translational invariance can have an anomaly.
In Section \ref{Symmetry}, we discuss the ambiguity 
from the viewpoint of the space-time symmetries,
such as chirality interchange, parity transformation,
charge conjugation and a 90$^\circ$ 
rotation, which should be satisfied by 
a fermion determinant in a general chiral gauge theory.
In Section \ref{Singular}, we consider gauge configurations
with a delta-function like singularity, for which
chiral determinants can have an ambiguity in its phase.
In Section \ref{anomalyfree}, 
we reconsider the anomaly for singular gauge transformations
in the 2D U(1) chiral gauge theory 
discovered in Ref. \cite{Anomaly}.
In Section \ref{generalization}
we generalize our results to any even dimension.
Section \ref{Summary} is devoted to summary and discussions.
 
\vspace*{1cm}

\section{Brief review of 2D U(1) chiral gauge theory 
and translational anomaly}
\label{Review}
\setcounter{equation}{0}

We consider a 2D U(1) chiral gauge theory.
The action is given by
\beq
S=- \int \dd ^2  x ~
 \bar{\psi } (x) \sigma_\mu (\del _\mu + i A_\mu (x) ) \psi (x) ,
\label{actionR}
\eeq
where 
$\sigma_1 = 1$ and $\sigma_2 = i$
and $\psi (x)$ is a two-dimensional right-handed
Weyl fermion
in a finite box $0 \le x_\mu < \ell$.
%So far we have been considering a right-handed Weyl fermion.
%Here we also consider a left-handed Weyl fermion, 
%whose action in the continuum
%is given by
%\beq
%S= - \int \dd ^2  x ~
% \bar{\psi _L} \sigma_\mu ^* (\del _\mu + i A_\mu ) \psi _L .
%\eeq
The boundary condition for the gauge field is taken to be 
periodic:
\beq
A_\mu (x + n \ell ) 
= A_\mu (x),
\eeq
while the one for the fermion is taken to be general : 
\beq
\psi (x + n \ell ) 
= - \ee ^{2 \pi i n_\mu b _\mu} \psi (x) ,
\label{fermionBC}
\eeq
where $n_\mu$ is an integer vector and
$b_\mu$ is a real phase.
The location of the boundary, 
on which we impose the boundary condition for the fermion, 
is irrelevant in vector-like gauge theories,
but it could become relevant in chiral gauge theories,
as we will see shortly.
We therefore assume throughout this paper that the boundary condition
for the fermion is imposed on $\{ x_1 = 0 \} \cup \{ x_2 = 0 \}$.
We denote the chiral determinant as
\beq
D(A_\mu , b _\mu ) =
\int {\cal D} \psi {\cal D} \bar{\psi} ~
\ee ^ {- S[A,\psi ]}  .
\label{pathintegral}
\eeq
The chiral determinant $D(A_\mu , b _\mu )$
can be exactly calculated in the continuum for finite $\ell$ \cite{Finite}.

%%%%%%%%%%%%%%%constant

We first restrict the boundary condition for the fermion
to be antiperiodic.
The continuum result for
a constant gauge background, namely for 
$A_\mu (x) = \alpha _\mu$(const.),
has been obtained in the context of string theory \cite{AMV}.
%The result can be stated in the following way.
%Without loss of generality, we can restrict ourselves
%to $ -\pi < b _\mu < \pi $.
The result can be expressed as
\beq
D(\alpha_\mu, 0)
= \hat{\vartheta} (h_\mu ),
\label{exactconst}
\eeq
where $h_\mu = \frac{\ell \alpha_\mu}{2 \pi}$.
$\hat{\vartheta}(h_\mu)$ is defined as
\beq
\hat{\vartheta} (h_\mu)
= \ee ^{ - \pi (h_2)^2 + i \pi h_1 h_2}
\frac{\vartheta (z,i)}{\eta (i)} ,
\eeq
where $z=h_1+i h_2$,
and $\vartheta(z,\tau)$ and $\eta(\tau)$ are
the theta function and eta function
defined by
\beqa
\vartheta (z,\tau) &=&
\sum_{\nu=-\infty}^{\infty}
\ee ^{i \pi \tau \nu^2 + 2 \pi i \nu z } ,  \\
\eta (\tau ) &=& 
\ee ^{\frac{\pi}{12} i \tau}
\prod _{\nu=1} ^{\infty}
(1 - \ee ^{2\pi i \tau \nu}).
\eeqa
%We introduce a function
%\beq
%\hat{b} (h_\mu) =
%\sum_{n=-\infty}^{\infty}
%\ee ^{- \pi (n+h_2)^2 + i 2 \pi n h_1 + i \pi h_1 h_2} .
%\eeq
%We write $b_\mu$ as
%$b_\mu = b ' _\mu + 2 \pi k_\mu$,
%where $-\pi < b ' _\mu  + 2 \pi k_\mu$,
(\ref{exactconst}) differs from the formula in Ref. \cite{AMV}
by a phase factor.
The freedom in defining $D(\alpha_\mu, 0)$ can be fixed
by requiring the quantity to have reasonable properties
under parity transformation, charge conjugation, 
a 90$^\circ$ rotation
and a gauge transformation \cite{torus}. 
%general

Let us next turn to a general gauge background.
A general 2D U(1) gauge field
can be decomposed as \cite{SachsWipf}
\beq
A_\mu (x)
= \epsilon _{\mu\nu} 
\left(\frac{\pi k}{\ell ^2}x_\nu + \del _\nu \phi (x) \right)
        + \alpha_\mu + \del_\mu \chi (x)  ,
\label{Adecompose}
\eeq
where $\phi(x)$ and $\chi(x)$ are real periodic functions of 
$x_\mu$, and $\alpha_\mu$ are real constants.
$k$ is an integer, which identifies the topological class.
When $k\neq 0$, the fermion has zero modes and
the determinant vanishes.
Therefore, we only need to consider $k=0$
in order to calculate the determinant.
Under the change of variables
\beqa
\psi(x) &=& \ee ^{- i \chi (x)} \ee^{- \phi (x)} \psi ' (x) , \\
\bar{\psi}(x) &=& \ee ^{ i \chi(x)} \ee^{ \phi (x)} \bar{\psi}' (x) ,
\eeqa
the action becomes
\beq
S=- \int \dd ^2  x ~
 \bar{\psi}  ' (x)  \sigma_\mu (\del _\mu + i \alpha_\mu ) \psi  ' (x) ,
\eeq
which means that 
the path integral over $\psi '$ and $\bar{\psi} '$ gives
the chiral determinant under the constant gauge background 
$D(\alpha_\mu, 0)$.
The Jacobian for the change of variables, however, is nontrivial,
since it needs regularization,
and can be obtained as
\beq
J = \exp \left [ \frac{1}{4 \pi}
\int \dd ^2 x ( \phi \del ^2 \phi + i \phi \del ^2 \chi  )
 \right] ,
\eeq
requiring that all the gauge breaking part 
is put in the parity odd part \cite{NN}.
Putting all these together, the chiral determinant for general 
$A_\mu(x)$ for an antiperiodic boundary condition can be written as
\beq
D(A_\mu, 0)
= \hat{\vartheta} (h_\mu )
  \exp \left [ \frac{1}{4 \pi}
\int \dd ^2 x ( \phi \del ^2 \phi + i \phi \del ^2 \chi  )
 \right] .
\label{exactanti}
\eeq

A generalization of the above result to an arbitrary boundary
condition parametrized by $b_\mu$ as in (\ref{fermionBC})
is not as trivial as it appears
and has not been fully examined in the literature.
We first do it by representing a change of the boundary condition
as a delta-function like singularity in the gauge field :
\beq
A'_\mu (x) = A_\mu (x) + 2\pi P [b _\mu ] 
\delta (x_\mu) .
\label{singcontbc}
\eeq
We have defined a projection function $P [t]$ by
\beq
P [t] = t -  \nu \mbox{~~~~~~~~~~~~~for~~}|t-\nu | < 1/2 ,
\eeq
where $\nu$ is an integer.
If we decompose $A_\mu (x)$ as in (\ref{Adecompose}),
the decomposition for $A'_\mu (x)$ can be obtained
by the replacements
\beqa
\alpha_\mu &\rightarrow & \alpha _\mu + \frac{
2 \pi P [b _\mu ] }{\ell}  , \\
\chi & \rightarrow & \chi - 
\sum_{\mu = 1} ^{2} \frac{ 2 \pi P [b _\mu ] x_\mu }{\ell} .
\eeqa
Thus we obtain
\beqa
D(A' _\mu,0 ) 
%&=& \ee ^{i \eta} D \left(A_\mu+\frac{
%P_{2\pi} [b _\mu ] }{\ell}, 0 \right)  \\
&=& \ee ^{i \eta} 
\hat{\vartheta} (h_\mu + P [b_\mu] )
  \exp \left [ \frac{1}{4 \pi}
\int \dd ^2 x ( \phi \del ^2 \phi + i \phi \del ^2 \chi  )
 \right] ,
\label{exactgeneral_abc}
\\
\eta &=& - \frac{1}{2}
\left[ 
P [b _1]   \left. \int \dd x_2 \del _1 \phi \right|_{x_1 = 0}
+  P [b _2 ] 
\left. \int \dd x_1 \del _2  \phi \right| _{x_2 = 0} \right] .
\eeqa
This result, however, breaks the translational invariance.
One can see from the above derivation that
the breaking of the invariance comes from the gauge dependence
(or $\chi$-dependence) of the expression (\ref{exactanti}),
and as a consequence it lies only in the phase of the 
fermion determinant.
Note that the formal expression (\ref{pathintegral})
for the fermion determinant in the continuum
in terms of path integral
has the translational invariance as well as the gauge invariance
for any boundary conditions $b _\mu$
for the chiral fermion being considered.
In this sense, this should be called a translational anomaly.
If one considers vector-like gauge theories, 
the translational anomaly exactly cancels,
unless one takes different
boundary conditions for the left-handed and
right-handed components of the fermion.
However, this is not necessarily the case when one considers
anomaly-free chiral gauge theories.
Boundary conditions should satisfy a certain condition in order to
make the whole system translationally invariant.
We will give the condition explicitly
in Section \ref{anomalyfree}.
The translational anomaly is a notion which 
is definitely independent of the gauge anomaly in this sense.

On the other hand, 
one can cancel the translational anomaly
by adding the local counterterm
\beq
\delta =   \frac{1}{2}
\left[ 
 P [b _1]   \left. \int \dd x_2  A_2 \right|_{x_1 = 0}
-  P [b _2 ] 
\left. \int \dd x_1 A_1  \right| _{x_2 = 0} \right] ,
\eeq 
without spoiling the properties of the fermion determinant
under parity transformation, charge conjugation, 
a 90$^\circ$ rotation and a gauge transformation,
which have been used to fix the phase of 
$D(\alpha_\mu, 0)$ in (\ref{exactconst}).

Thus by imposing the translational invariance, we obtain
\beqa
D(A_\mu, b_\mu) &=&  \ee^{- i \delta}  D(A' _\mu,0 )  \n
&=& \hat{\vartheta} (h_\mu + P [b_\mu] )
  \exp \left [ \frac{1}{4 \pi}
\int \dd ^2 x ( \phi \del ^2 \phi + i \phi \del ^2 \chi  )
 \right]
\ee ^{i \pi (h _ 1 P [b _ 2] - h _2 P [b _1])  } .
\label{exactgeneral_m}
\eeqa
Note, however, 
that $D(A_\mu, b_\mu)$ differs from $D(A' _\mu,0 )$
by a phase factor.
This shows that the requirement on the chiral determinant
that a delta-function like singularity in the gauge field
can be absorbed by a change of the boundary condition for the fermions,
is not compatible with translational invariance.
We will discuss this issue in a more general setup
in Section \ref{BC}.
%where we also 
%show that the overlap formalism realizes this feature manifestly
%on the lattice.
 
\vspace*{1cm}

\section{Overlap formalism and the Wigner-Brillouin phase choice}
\label{Overlap}
\setcounter{equation}{0}

In this section,
we review the overlap formalism \cite{NN}.
Throughout this paper, we use a simplified version
first given in Ref. \cite{KikukawaNeuberger}, but all the results
below would equally apply to the original version.
We describe the formalism for the 2D U(1) case we are considering,
but generalization to other cases are straightforward.
We introduce a two-dimensional lattice
$\Lambda_L = \{ ( n_1,n_2) ~ | ~
n_\mu \in {\cal Z},~0 \le n_\mu < L,~\mu=1,2 \}$.
Denoting the lattice spacing by $a$,
the physical extent of the lattice is given by
$\ell = a L$, which should be fixed when we take the continuum
limit $a \rightarrow 0$.
%In what follows, we take $\ell$ to be unity.

We consider a many-body Hamiltonian
\beq
{\cal H} (U_{n\mu},b_\mu)
= \sum_{n_\mu \in \Lambda_L} \sum_{m_\mu \in \Lambda_L}
\left(
\begin{array}{cc}
\alpha_n ^{\dagger} & \beta_n ^\dagger
\end{array}
\right)
\left(
\begin{array}{cc}
B_{nm} - M \delta_{nm} &  C_{nm} \\
C_{nm} ^\dagger  & - (B_{nm} - M \delta_{nm})
\end{array}
\right)
\left(
\begin{array}{c}
\alpha_m  \\
\beta_m
\end{array}
\right),
\label{eq:mbham}
\eeq
where
\beqa
C_{nm} &=& \frac{1}{2} \sum_{\mu=1}^2 \sigma_\mu 
(\delta _{m,n+\hat{\mu}} ^{(b)} U_{n \mu}
- \delta _{n,m+\hat{\mu}} ^{(-b)} U_{m \mu}^{\dagger} ),  \\
B_{nm} &=& \frac{1}{2} \sum_{\mu=1}^2 
( 2 \delta_{nm} - 
\delta _{m,n+\hat{\mu}} ^{(b)} U_{n \mu}
- \delta _{n,m+\hat{\mu}} ^{(- b)} U_{m \mu}^{\dagger}) .
\eeqa
$ \delta _{m,n+\hat{\mu}} ^{(b)} $ is defined by
\beq
\delta _{m,n+\hat{\mu}} ^{(b)} =
\delta_{m,n+\hat{\mu}} - \delta_{m+(L-1)\hat{\mu},n} \ee ^{ 2 \pi i b _\mu} ,
\eeq
where the second term is there to
ensure the boundary condition for the fermions
which corresponds to (\ref{fermionBC}).
$\alpha_n$ and $\beta_n$ are fermionic operators 
which obey the canonical anticommutation relations:
\beqa
\{ \alpha_n , \alpha _m ^\dagger \} &=& \delta_{n,m} , \\
\{ \beta_n , \beta _m ^\dagger \} &=& \delta_{n,m},
\eeqa
and zero for the rest of the anticommutators.
$M$ is a mass parameter which satisfies $0<M<1$
and should be kept fixed when we take the continuum limit.
We denote the ground state of the many-body Hamiltonian 
${\cal H}(U_{n\mu},b_\mu)$ as 
$| 0 \rangle _{U,b} $.

We consider yet another Hamiltonian
${\cal H}^{\infty}$
which can be obtained by formally 
taking the limit $M \rightarrow - \infty$
of ${\cal H}(U_{n\mu},b_\mu)/|M|$.
Explicitly, ${\cal H}^{\infty}$ can be written as
\beq
{\cal H} ^{\infty}
= \sum_{n_\mu \in \Lambda_L} 
(\alpha_n ^\dagger \alpha_n  
- \beta_n ^\dagger \beta_n),
\label{eq:haminf}
\eeq
and the ground state $|0\rangle$ of this Hamiltonian can be given as
\beq
| 0 \rangle = \prod_{n _\mu \in \Lambda_L}
\beta _n ^\dagger |v\rangle ,
\eeq
where $|v\rangle$ is a kinematic vacuum defined as a state
which is annihilated by all of $\alpha _{n}$ and $\beta _n$.
The order of the product $\prod_{n_\mu \in \Lambda_L}$
should be specified as one wishes.

Now the basic idea of the overlap formalism
is to define a lattice-regularized fermion determinant
by the overlap ``$\langle 0 | 0 \rangle _{U,b}$'',
where we have put inversed commas,
%``~'', 
since the expression
is not complete in the sense that it is defined only 
up to a phase factor.
We have to fix the $U_{n\mu}$ dependence of the phase factor
of the state $| 0 \rangle _{U,b}$.
This can be done by requiring that
\beq
~_{U=1,b ^{(r)}} \langle 0 |
0 \rangle_{U,b} 
\eeq
should be real positive, where $b ^{(r)}_\mu$ is taken
independently of $U_{n\mu}$.
This is referred to as the Wigner-Brillouin phase choice
and the ground state of ${\cal H}(U_{n\mu},b_\mu)$ 
which satisfies the
above condition is denoted 
by $| 0 \rangle _{U,b} ^{\wb}$.
The state $|0 \rangle_{U=1,b^{(r)}}$ is called the reference
state.
Here we note that there is an ambiguity in the choice of 
$b_\mu ^{(r)}$, which we discuss below.
Let us denote the lattice-regularized fermion determinant
defined by the overlap formalism as
\beq
D_{lat} (U_{n\mu}, b_\mu; b_\mu ^{(r)}) 
= \langle 0 | 0 \rangle _{U,b} ^{\wb} ,
\label{eq:overlapdef}
\eeq
where the Wigner-Brillouin phase choice has been
taken with the reference state obeying the boundary condition
given by $b ^{(r)} _\mu$.
One of the most important features of the overlap formalism
is that the violation of the gauge invariance
resides only in the phase of the fermion determinant \cite{NN}.

\vspace*{1cm}

\section{Continuum limit of the overlap formalism for
arbitrary boundary conditions}
\label{BC}
\setcounter{equation}{0}

In this section, we study the continuum limit
of the overlap formalism.
The quantity we are interested in is given by
\beq
\frac{
\lim_{a \rightarrow 0} 
D_{lat}(U_{n\mu},b_\mu;b_\mu ^{(r)} )
}{
\lim_{a \rightarrow 0} 
D_{lat}(1,b_\mu;b _\mu ^{(r)})} \  ,
\label{quantity}
\eeq
where 
\beq
U_{n \mu} = \exp \left[ ~ i a   \int_{0} ^{1} 
A_\mu (a(n+ t \hat{\mu})) ~ \dd t ~ \right] .
\label{eq:linkvar}
\eeq
By taking the ratio of the fermion determinants,
we have dropped the irrelevant constant factor
independent of $U_{n\mu}$.
In all the figures in this paper,
we plot fermion determinants with this normalization.
We assume that $A_\mu (x)$ has no delta-function like singularities,
and therefore that all the link variables $U_{n\mu}$ go to unity 
in the continuum limit.
Gauge configurations with delta-function like singularities will
be considered in Section \ref{Singular} and \ref{anomalyfree}.

We first consider an invariance of the fermion determinant
under translations of the background gauge configuration.
Let us define a shifted gauge configuration by
\beq
U'_{n\mu} = U_{n' \mu} ,
\eeq
where $n' _\mu \equiv n_\mu - s_\mu$ (mod $L$) and $s_\mu$
is an integer vector.
% \in \Lambda_L$.
Then we have
\beq
D_{lat}(U_{n\mu},b _\mu;b^{(r)}_\mu=b _\mu)=
D_{lat}(U'_{n\mu},b _\mu;b^{(r)}_\mu=b _\mu) .
\label{latticetranslinv}
\eeq
Thus if we take $b ^{(r)}_\mu=b_\mu $,
we have manifest translational invariance.
On the other hand,
if we take $b ^{(r)} _\mu \neq b _\mu$,
the translational invariance is broken on the lattice.

Therefore, it is natural to expect that
in order to reproduce the exact result
(\ref{exactgeneral_m})
obtained in the continuum 
by imposing the translational invariance,
we have to take $b ^{(r)}_\mu=b_\mu $.
Explicitly, we expect
\beq
\frac{
\lim_{a \rightarrow 0} 
D_{lat}(U_{n\mu},b_\mu;b_\mu ^{(r)} =b_\mu )
}{
\lim_{a \rightarrow 0} 
D_{lat}(1,b_\mu;b _\mu ^{(r)} = b_\mu )}
=
\frac{D(A_\mu,b_\mu)}{D(0,b_\mu  )}  \ ,
\label{eq:statement}
\eeq
where $D(A_\mu,b_\mu)$ is given by (\ref{exactgeneral_m}).
Checks of this statement
have been done for $b_\mu = 0$
both numerically \cite{NN,torus,Anomaly} 
and analytically \cite{FRD,Fosco}.
We have checked analytically that the statement
(\ref{eq:statement}) holds for constant gauge backgrounds
by generalizing the analysis made in Ref. \cite{Fosco}
for $b _\mu = 0$ to an arbitrary $b_\mu$.

We can further ask what we get for the quantity (\ref{quantity})
if we take $b_\mu ^{(r)} \neq b_\mu$.
We first note that the overlap determinant satisfies the 
property
\beq
D_{lat} (U _{n\mu} '' , b _\mu ; 
b _\mu ^{(r)}  ) 
= D_{lat} (U_{n\mu} , b _\mu + d _\mu; 
b _\mu ^{(r)}  ) ,
\label{eq:absorblat}
\eeq
where $U _{n\mu} '' $ is defined by
\beq
U _{n\mu} '' = 
\left\{
\begin{array}{ll}
\ee^{2 \pi i d _\mu} U_{n\mu}  & \mbox{for}~~~n_\mu = L-1  \\
U_{n\mu}  & \mbox{otherwise} , 
\end{array}
\right.
\eeq
and $d_\mu$ is a real phase.
It is therefore natural to define the continuum counterpart of
$D_{lat}(U_{n\mu},b_\mu;b_\mu ^{(r)} )$
by the relation
\beq
D(A_\mu , b_\mu ; b_\mu ^{(r)})
= D(A_\mu '' , b_\mu ^{(r)})   ,
\label{contcounterpart}
\eeq
where
\beq
A''_\mu (x) = A_\mu (x) + 2 \pi P [b _\mu - b _\mu ^{(r)} ] 
\delta (x_\mu) .
\label{Adoubleprime}
\eeq
Note that we have $D(A_\mu , b_\mu ; b_\mu ^{(r)} = b_\mu)
= D(A_\mu  , b_\mu ) $.
Using the procedure that led to (\ref{exactgeneral_abc}),
we obtain
\beq
D(A_\mu, b_\mu;b_\mu ^{(r)})
= \ee ^{i \beta} \ee ^{i \gamma} 
D(A_\mu, b_\mu) ,
\label{exactgeneral_r}
\eeq
where
\beqa
\beta &=& \frac{1}{2}
\left[ 
 P [b _1 - b _1 ^{(r)} ]  
 \left. \int \dd x_2  A_2 \right|_{x_1 = 0}
- 
 P [b _2 - b _2  ^{(r)} ]  
\left. \int \dd x_1 A_1  \right| _{x_2 = 0} \right] ,
\label{beta}
\\
\gamma &=& \pi
\{ P[b_1 ^{(r)}] P[b_2] 
+ P[b_1 - b_1 ^{(r)}] (  P[b_2]  + P[b_2 ^{(r)}] ) \}
\n
&~& 
- \pi \{ P[b_2 ^{(r)}] P[b_1] 
+ P[b_2 - b_2 ^{(r)}] (  P[b_1]  + P[b_1 ^{(r)}] ) \}  .
\label{gamma}
\eeqa
We have used the identity
\beq
\hat{\vartheta}(h_\mu + n_\mu) =
\ee^{i \pi (n_1 h_2 - n_2 h_1)} \hat{\vartheta} (h_\mu)  .
\label{iden}
\eeq
What we expect to hold is 
\beq
\frac{
\lim_{a \rightarrow 0} 
D_{lat}(U_{n\mu},b_\mu;b_\mu ^{(r)} )
}{
\lim_{a \rightarrow 0} 
D_{lat}(1,b_\mu;b _\mu ^{(r)})}
=
\frac{D(A_\mu,b_\mu;b_\mu ^{(r)})}
{D(0,b_\mu  ;b_\mu ^{(r)})}   \ .
\label{eq:statement_r}
\eeq
For $b_\mu ^{(r)} = b_\mu $, 
(\ref{eq:statement_r}) reduces to
(\ref{eq:statement}).
Note that the phase $\ee ^{i \gamma}$ in (\ref{exactgeneral_r}), 
which is independent of $A_\mu (x)$,
cancels between the numerator and the denominator
in the r.h.s. of eq. (\ref{eq:statement_r}).
Therefore, the effect of having $b_\mu ^{(r)}$ different
from $b_\mu $ is essentially given by the phase factor 
$\ee ^{i \beta}$, where $\beta$ is proportional to the
Polyakov loops along the boundaries.
This gives rise to a translational anomaly.

\begin{figure}
\begin{center}
  \leavevmode
  \epsfxsize=8cm
  \epsfbox{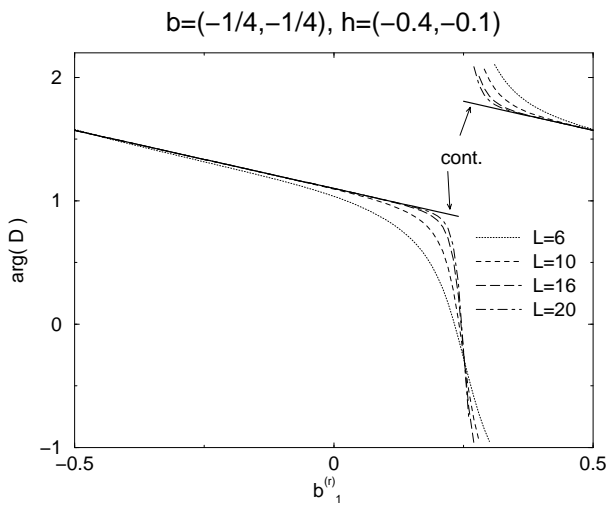}
\end{center}
\caption{
The argument of the normalized overlap chiral determinants 
for a constant gauge background
$U_{n\mu}=\ee ^{2\pi i h_\mu/L}$ with $h_\mu = (-0.4,-0.1)$ and
$b_\mu = (-1/4,-1/4)$ is plotted against
$b_1 ^{(r)}$, where we take $b_1 ^{(r)}=b_2 ^{(r)}$.
The bold solid line represents
the continuum prediction (\ref{exactgeneral_r})
for $A_\mu (x) = 2 \pi h_\mu / \ell$.
%$\hat\vartheta(h_\mu+b_\mu)/\hat\vartheta(b_\mu)$. 
}
\label{fig:b0_Udet3}
\end{figure}

In Fig. \ref{fig:b0_Udet3}
we plot the argument of the normalized overlap chiral determinants 
for a constant gauge background
$U_{n \mu} = e^{i2\pi h_\mu/L}$ with $h_\mu = (-0.4,-0.1)$ and
$b_\mu = (-1/4,-1/4)$ as a function of 
$b_1 ^{(r)}$, where we take $b_1 ^{(r)}=b_2 ^{(r)}$.
We can see that the statement (\ref{eq:statement_r}) holds.
Note that (\ref{beta}) gives rise to a
discontinuity in the phase of the determinant
at $b_1 ^{(r)} =1/4 $ and at $b_2 ^{(r)} =1/4 $.
Accordingly, the convergence to the continuum limit is slower near 
$b_1 ^{(r)} =1/4 $ in Fig. \ref{fig:b0_Udet3}.
In what follows, we examine
the statement (\ref{eq:statement_r}) in more detail
for $b_\mu ^{(r)}=b_\mu$ (the translationally invariant case) 
and $b_\mu ^{(r)}=0$ 
as a typical example for a translationally non-invariant case.

\begin{figure}
\begin{center}
  \leavevmode
  \epsfxsize=8cm
  \epsfbox{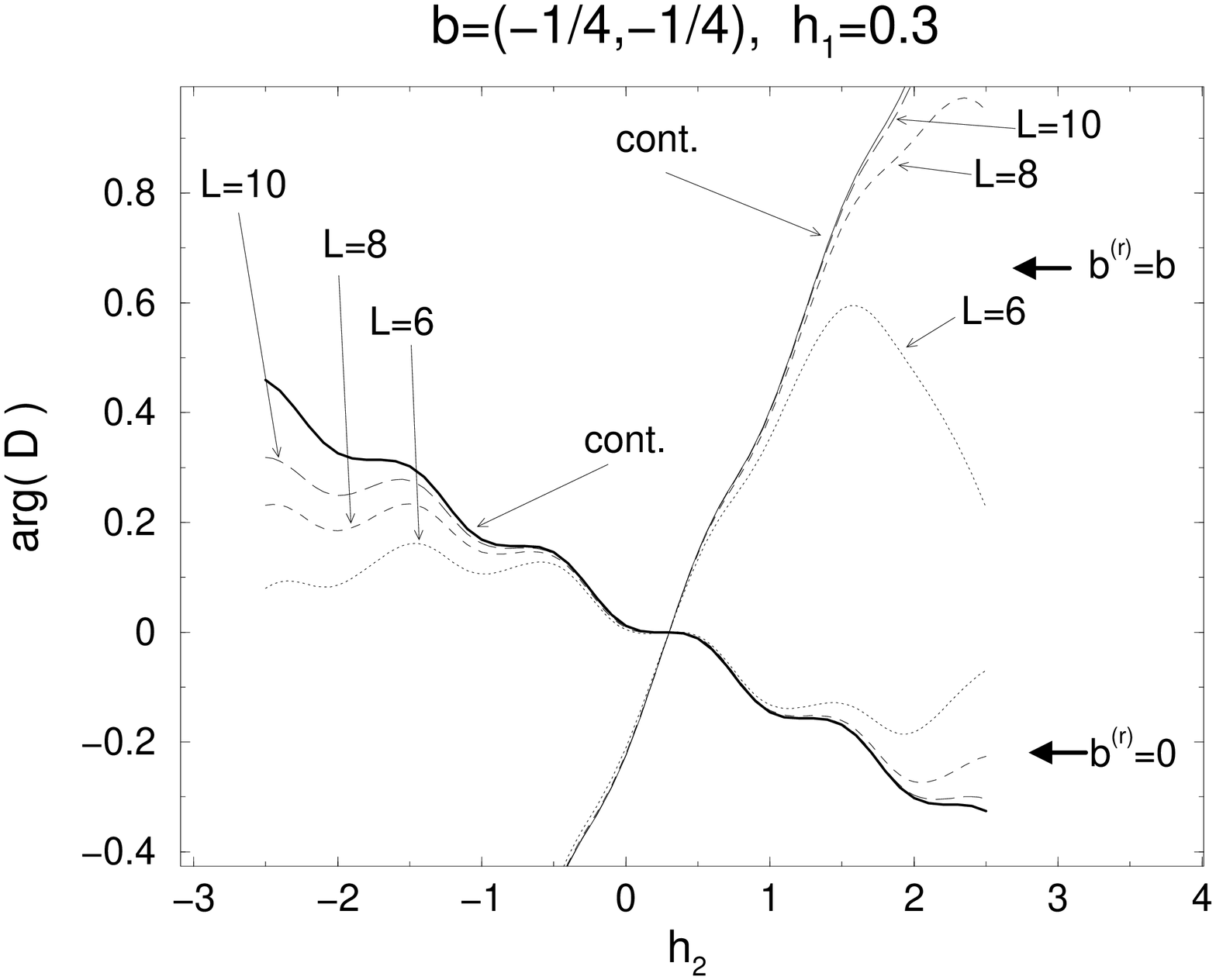}
\end{center}
\caption{
The argument of the normalized overlap determinant for a constant 
gauge background $U_{n\mu}=\ee ^{2\pi i h_\mu/L}$ with a boundary condition
$b_\mu = (-1/4,-1/4)$ is plotted as a function of $h_2$
for $h_1 =0.3$.
The boundary condition for the reference state
is taken to be either $b_\mu ^{(r)}=0$
or $b_\mu ^{(r)}=b_\mu$.
The continuum results (\ref{exactgeneral_r})
for $A_\mu (x) = 2 \pi h_\mu / \ell$
are shown by the bold solid lines for the corresponding $b_\mu ^{(r)}$ .}
\label{fig:h2_Udet}
\end{figure}

Let us first consider a constant gauge background
parametrized by $h_\mu$.
We first fix the boundary condition as $b_\mu=(-1/4,-1/4)$
and examine the $h_\mu$ dependence of the fermion determinant.
In Fig. \ref{fig:h2_Udet}
we plot the argument of the normalized fermion determinant 
against $h_2$ for $h _1  = 0.3$.
The boundary condition for the reference state is taken
to be either $b^{(r)}_\mu=0$ or $b^{(r)}_\mu=b_\mu$.
%$L$ is taken to be $L=6,10,16,20$.
%The continuum results (\ref{exactgeneral_r})
%are shown by the bold solid lines for the corresponding
%$b_\mu^{(r)}$.
We see that the data for both $b^{(r)}_\mu$
%$b^{(r)}_\mu=b_\mu$ 
seem to converge 
to the corresponding continuum results as we increase $L$,
although finite lattice spacing effects increase for larger $|h_2|$ 
as expected.
%As we already mentioned,
%we can obtain the continuum limit ($L=\infty$)
%of the overlap determinant analytically
%for $b^{(r)}_\mu=b_\mu$, 
%and show that it agrees indeed with the expected continuum result.

\begin{figure}
\begin{center}
  \leavevmode
  \epsfxsize=8cm
  \epsfbox{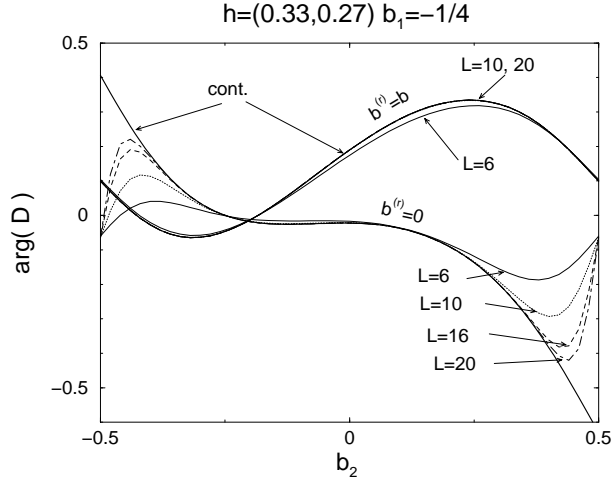}
\end{center}
\caption{
The argument of the normalized overlap determinant for a
constant gauge background
$U_{n\mu}=\ee ^{2\pi i h_\mu/L}$ with 
$h_\mu = (0.33,0.27)$ is plotted 
as a function of $b_2$ for $b_1=-1/4$.
The boundary condition for the reference state
is taken to be either $b_\mu ^{(r)}=0$
or $b_\mu ^{(r)}=b_\mu$.
The continuum results
(\ref{exactgeneral_r})
for $A_\mu (x) = 2 \pi h_\mu / \ell$ 
are shown by the bold solid lines
for the corresponding
$b_\mu ^{(r)}$.
The results for $b_\mu ^{(r)}=b_\mu$ with $L\geq 10$
cannot be distinguished from the corresponding 
continuum result in this figure.
}
\label{fig:b_Udet2}
\end{figure}

We next examine the $b_\mu$ dependence of the fermion determinant
for a fixed constant gauge background.
In Fig. \ref{fig:b_Udet2}
we take $h_\mu=(0.33,0.27)$ and plot
the argument of the normalized overlap determinant 
against $b_2$ for $b_1=-1/4$.
The boundary condition for the reference state
is taken to be either $b_\mu ^{(r)}=0$
or $b_\mu ^{(r)}=b_\mu$.
%The continuum results (\ref{exactgeneral_r}) for the corresponding
%$b_\mu ^{(r)}$ are shown by the bold solid lines.
%$L$ is taken to be $6,10,16,18,20$.
The data for both $b_\mu ^{(r)}$ seem to converge to the 
corresponding continuum results. 
%Note again that we have obtained
%the continuum limit ($L=\infty$)
%of the overlap determinant analytically
%for $b^{(r)}_\mu=b_\mu$, 
%which agrees indeed with the expected continuum result.
The continuum result for $b_\mu ^{(r)}=0$
is discontinuous at $b_2 = \pm 1/2$.
%,which corresponds to the periodic boundary condition.
Accordingly, the convergence to the continuum limit becomes
slower as $b_2$ gets closer to $\pm 1/2$.
The continuum result for $b_\mu ^{(r)}=b_\mu$, on the other hand,
is a continuous function of $b_\mu$ and the data converge to 
the continuum result rapidly for all $b_2$.

\begin{figure}
\begin{center}
  \leavevmode
  \epsfxsize=8cm
  \epsfbox{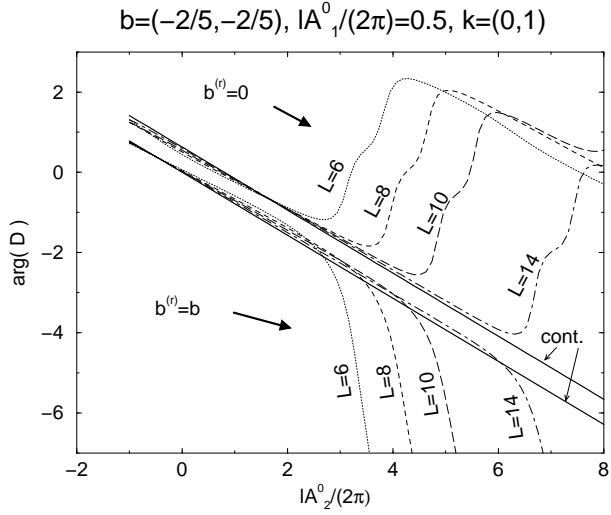}
\end{center}
\caption{
The argument of the normalized 
overlap determinant for a sine-type gauge
configuration (\ref{sinetype})
with $k_\mu =(0,1)$ and $b_\mu = (-2/5,-2/5)$
is plotted against $\ell A^0_2/(2\pi)$ 
for $\ell A^0_1=\pi$.
The boundary condition for the reference state
is taken to be either $b_\mu ^{(r)}=0$
or $b_\mu ^{(r)}=b_\mu$.
The continuum results (\ref{exactgeneral_r}) for the gauge 
configuration  $A_\mu=A^0_\mu\cos(2\pi k \cdot x /\ell)$ are shown
by the bold solid lines for the corresponding $b_\mu ^{(r)}$.
}
\label{fig:cos}
\end{figure}

We next consider a more general gauge configuration.
In Ref. \cite{NN}, a sine-type gauge configuration has been 
considered for $b_\mu = b_\mu ^{(r)}=0$
and the result showed a good agreement with the continuum result
(\ref{exactanti}).
In Fig. \ref{fig:cos} we plot 
the argument of the normalized 
overlap determinant for a sine-type gauge
configuration 
\beq
U_{n \mu}=\exp \left[i \frac{\ell A^0_\mu }{L}
 \cos \left( \frac{ 2\pi k \cdot n +\pi k_\mu }{L} \right) \right] ,
\label{sinetype}
\eeq
with $k_\mu = (0,1)$ and $b_\mu = (-2/5,-2/5)$,
against $\ell A^0_2/(2\pi)$
for a fixed $\ell A^0_1 = \pi$.
The boundary condition for the reference state is
taken to be either $b_\mu ^{(r)}=0$
or $b_\mu ^{(r)}=b_\mu$.
The data for both $b^{(r)}_\mu$ 
%and $b_\mu ^{(r)} = b_\mu$
seem to converge to the corresponding continuum results,
% (\ref{exactgeneral_r}),
although finite lattice spacing effects 
increase for larger $|\ell A^0_2|$, as expected.

%chiral determinant for any boundary conditions can be reproduced
%correctly if we always adopt the antiperiodic 
%boundary condition for the reference state
%irrespective of the boundary conditions for the fermion
%under consideration.
%$b ^{(r)} =0$ is essential in reproducing the continuum result.
%for general $b _\mu$.

\begin{figure}
\begin{center}
  \leavevmode
  \epsfxsize=8cm
  \epsfbox{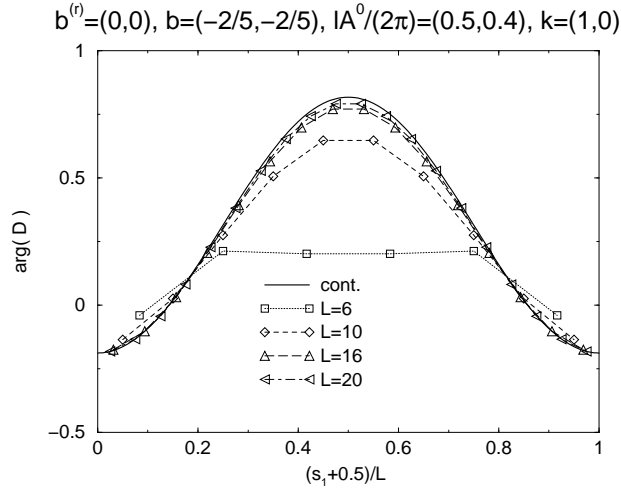}
\end{center}
\caption{
The argument of the overlap determinant 
for shifted gauge configuration $U'_{n\mu}=U_{n'\mu}$,
$n' _\mu \equiv n_\mu - s_\mu$ 
is plotted against the shift on physical scale 
$(s_1+1/2)/L$,
where the shift is taken to be symmetric in the two directions 
($s_1 = s_2$).
We take the original configuration
to be a sine-type (\ref{sinetype}),
%$U_{n \mu}=\exp[i a A^0_\mu \cos(2\pi n_2/L+\pi/L)]$,
where $\ell A_\mu ^0/(2\pi) = (0.5,0.4)$ and $k_\mu =(1,0)$.
The boundary condition is taken to be $b_\mu = (-2/5,-2/5)$.
For the reference state, it is taken to be
antiperiodic ($b^{(r)}_\mu=0$).
The bold solid line represents the continuum result
(\ref{exactgeneral_r}).
}
\label{fig:TransOsc2}
\end{figure}

Let us see explicitly how the overlap formalism for
$b _\mu ^{(r)} \neq b_\mu$ reproduces 
the expected translational anomaly.
In Fig. \ref{fig:TransOsc2}
we plot the argument of the normalized overlap determinant 
for a shifted gauge configuration $U'_{n\mu}=U_{n'\mu}$,
$n'_\mu \equiv n_\mu - s_\mu$
against the shift on physical scale $(s_1 + 1/2) /L$,
where the shift is taken to be symmetric in the two directions $s_1=s_2$.
The original configuration 
is taken to be a sine-type
(\ref{sinetype}) with $k_\mu = (1,0)$, $\ell A_\mu ^0/(2\pi) = (0.5,0.4)$
and $b_\mu = (-2/5,-2/5)$.
We take $b_\mu ^{(r)} = 0$.
One can see that
the translational anomaly given by (\ref{exactgeneral_r})
is clearly reproduced by the overlap formalism
in the continuum limit.

Finally, we check (\ref{eq:statement_r}) for a more generic case.
We take the gauge configuration to be
\beq
U_{n\mu}=\ee ^ { 2\pi i h_\mu/ L}
\exp \left[  i \frac{\ell A_\mu ^0}{L}
\cos \left( \frac{2\pi k \cdot(n- s)+\pi k_\mu}{L} \right) \right] ,
\label{generic}
\eeq 
where $h_\mu=(0.43,0.13)$, $k_\mu=(0,1)$ and $s_\mu /L=(0.2,0.6)$.
The boundary conditions are taken to be
$b_\mu=(-0.4,-0.15)$ and $b_\mu ^{(r)}=(0.225,-0.05)$.
We plot the normalized overlap determinant 
against $\ell A_2 ^0/(2\pi)$ for $\ell A_1 ^0/(2\pi)=0.37$. 
The data are seen to converge to the continuum prediction
(\ref{exactgeneral_r}).

\begin{figure}
\begin{center}
  \leavevmode
  \epsfxsize=8cm
  \epsfbox{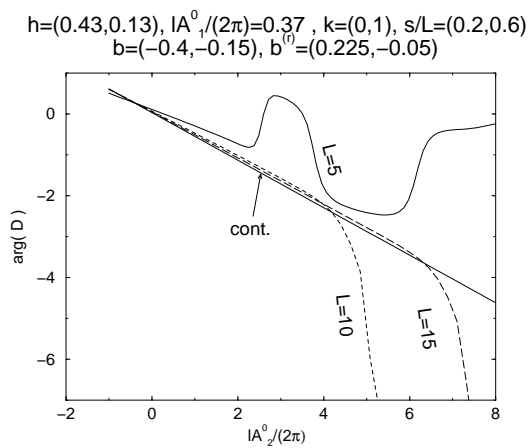}
\end{center}
\caption{
The argument of the normalized overlap determinant for a 
gauge configuration (\ref{generic})
with $h_\mu=(0.43,0.13)$, $k_\mu=(0,1)$ and $s_\mu/L=(0.2,0.6)$
is plotted against $\ell A_2 ^0/(2\pi)$ for $\ell A_1 ^0/(2\pi)=0.37$. 
The boundary conditions are taken to be
$b_\mu=(-0.4,-0.15)$ and $b_\mu ^{(r)}=(0.225,-0.05)$.
The continuum result (\ref{exactgeneral_r})
%for the gauge configuration 
%$A_\mu=A_\mu ^0 \cos (2\pi x_2/\ell) + 2\pi h_\mu/\ell$ 
is shown by the bold solid line.
}
\label{fig:general}
\end{figure}

Having confirmed that the continuum limit of the overlap
formalism gives (\ref{eq:statement_r}),
let us discuss the physical implications of this result.
We have seen that the ambiguity of the overlap formalism,
which lies in the choice of the boundary condition for the reference state,
corresponds to the phase ambiguity
of the chiral determinant on a two-dimensional torus
proportional to the Polyakov loops along the boundaries.
This gives rise to a translational anomaly in general.
The identity (\ref{latticetranslinv}) 
shows that translational invariance can be preserved
by taking $b_\mu ^{(r)} = b_\mu$.
On the other hand, 
the identity (\ref{eq:absorblat}) shows that, 
when one absorbs a delta-function like 
singularity in the gauge field
by a change of the boundary condition for the fermion,
$b_\mu ^{(r)}$ should be kept fixed.
Therefore, the requirement on the fermion determinant
that a delta-function like 
singularity in the gauge field can be absorbed by a change of
the boundary condition for the fermion, is not compatible
with translational invariance.
Note that this feature is not restricted to the 2D U(1) case,
since (\ref{latticetranslinv}) and (\ref{eq:absorblat})
hold for general chiral gauge theories on a torus.
%This will be of great importance when we discuss
%the anomaly under singular gauge transformations in Section
%\ref{anomalyfree}.
 
\vspace*{1cm}

\section{Symmetries of the fermion determinant}
\label{Symmetry}
\setcounter{equation}{0}

In this section, we examine the ambiguity of $b _\mu ^{(r)}$
from the viewpoint of 
symmetries that the fermion determinant should possess
for a general chiral gauge theory.
We consider the space-time symmetries,
such as chirality interchange, parity transformation,
charge conjugation and a 90$^\circ$ rotation,
which have been discussed in Ref. \cite{NN}
in the infinite volume.
What we do here is to repeat their argument for
a finite lattice with arbitrary boundary conditions
for fermions.

Here we need to consider 
Weyl fermions with the opposite chirality.
%which we denote by
%$D^{R}(A,b)$ and $D^{L}(A,b)$, respectively.
%In the continuum, we have the following relation.
%\beq
%D^{R}(A,b)=D^{L}(A,b ')^* ,
%\label{LRexchage}
%\eeq
%where $b '$ is given by
In the overlap formalism,
the fermion determinant for a left-handed Weyl fermion,
namely with the opposite chirality to the one we have
been considering, can be obtained \cite{NN}
by simply flipping the sign of the many-body Hamiltonians
(\ref{eq:mbham}) and (\ref{eq:haminf}) for the right-handed
Weyl fermion.
We denote the fermion determinant for each chirality
defined within the overlap formalism by
$D_{lat}^{R}$ and $D_{lat}^{L}$, respectively.
Note that 
$D_{lat}^{R}(U_{n\mu},b_\mu;b_\mu ^{(r)})$ and
$D_{lat}^{L}(U_{n\mu},b_\mu;b_\mu ^{(r)})$
have a phase ambiguity independent of both $U_{n\mu}$ and
$b_\mu$ for each $b_\mu ^{(r)}$.
Here we assume that this residual phase ambiguity
has been fixed 
by requiring that the fermion determinant be real positive
for $U_{n\mu}=1$ and $b_\mu = b_\mu ^{(r)}$.
Then the following statements hold for any chiral gauge theory
in any even dimension $D$.

\noindent \underline{(i)~chirality interchange}
%The relation 
%between the determinants for opposite handedness}

The relation between $D_{lat}^{R}$ and $D_{lat}^{L}$
is given as
\beq
D_{lat}^{R}(U_{n\mu},b _\mu;b^{(r)}_\mu)=
D_{lat}^{L}(U_{n\mu},b _\mu ;b^{(r)}_\mu)^* .
\label{chiralityint}
\eeq

\noindent \underline{(ii)~parity transformation}

We consider the parity transformation of the gauge field:
\beqa
U ^P _{n\mu} =
\left\{
\begin{array}{ll}
U_{n^P,-\mu}
& \mbox{for}~~~\mu = 1,...,D-1  \\
U_{n^P, \mu} & \mbox{for}~~~\mu = D  ,
\end{array}
\right.
\label{Uparity}
\eeqa
where $n^P_\mu$ is defined by
\beq
n^P_\mu =
\left\{
\begin{array}{ll}
(L-1) - n_\mu
& \mbox{for}~~~\mu = 1,...,D-1  \\
n_\mu  & \mbox{for}~~~\mu = D  .
\end{array}
\right.   
\label{nparity}
\eeq
$U_{n,-\mu}$ is defined as usual by the hermitian conjugate
of the link variable residing on a link which stems from
the site $n_\mu$ to the $-\mu$ direction and can be given
explicitly as
\beq
U_{n,-\mu} =
\left\{
\begin{array}{ll}
U_{n-\hat{\mu},\mu}^\dagger
& \mbox{for}~~~~~n_\mu = 1,...,L-1  \\
U_{n+(L-1)\hat{\mu},\mu}^\dagger
& \mbox{for}~~~~~n_\mu = 0  .
\end{array}
\right.
\label{Uminus}
\eeq
Then, we have
\beq
D_{lat}^{R}(U_{n\mu} ,b _\mu;b^{(r)}_\mu)=
D_{lat}^{L}(U ^P _{n\mu},b ^P _\mu ;b^{(r)P} _\mu) ,
\label{parityinv}
\eeq
where $b ^P _\mu$ is given by
%(\ref{thetaparity}).
\beq
b_\mu ^P =
\left\{
\begin{array}{ll}
- b_{\mu}   & \mbox{for}~~~\mu = 1,...,D-1  \\
b_{\mu}  & \mbox{for}~~~\mu = D , \\
\end{array}
\right.
\label{thetaparity}
\eeq
and $b_\mu^{(r)P}$ is defined similarly.

\noindent \underline{(iii)~charge conjugation}

Under the charge conjugation of the gauge field,
we have
%\beq
%U ' _{\mu} (x) =U_{\mu}^*(x)   .
%\label{Uconjg}
%\eeq
%Then,
\beqa
D_{lat}^{R}(U_{n\mu},b _\mu;b^{(r)}_\mu)&=&
D_{lat}^{R}(U^* _{n\mu},-b_\mu ; - b^{(r)}_\mu) ~~
\mbox{for}~~~D=2 ,  \\
D_{lat}^{R}(U_{n\mu},b_\mu;b^{(r)}_\mu)&=&
D_{lat}^{L}(U^* _{n\mu},-b _\mu ; - b^{(r)} _\mu) ~~
\mbox{for}~~~D=4 ,  
\eeqa
and similar relations for $D \ge 6$.

\noindent \underline{(iv)~90$^\circ$ rotational invariance}

We consider a 90$^\circ$ rotation in the ($\alpha$,$\beta$) plane,
where $1 \le \alpha < \beta \le D$.
The gauge configuration is transformed as
\beqa
U ^{rot} _{n\mu} =
\left\{
\begin{array}{ll}
U_{n^{rot},\beta}
& \mbox{for}~~~\mu = \alpha  \\
U_{n^{rot},-\alpha}
& \mbox{for}~~~\mu = \beta  \\
U_{n^{rot}, \mu} & \mbox{otherwise},
\end{array}
\right.
\label{Urotation}
\eeqa
where $n^{rot}_\mu$ is defined by
\beq
n^{rot}_\mu =
\left\{
\begin{array}{ll}
n_\beta & \mbox{for}~~~\mu = \alpha \\
(L-1) - n_\alpha
& \mbox{for}~~~\mu = \beta  \\
n_\mu  & \mbox{otherwise} ,
\end{array}
\right.   
\label{nrotation}
\eeq
and $U_{n,-\mu}$ is defined by (\ref{Uminus}).
Then, we have
\beq
D_{lat}^{R}(U_{n\mu} ,b _\mu;b^{(r)}_\mu)=
D_{lat}^{R}(U ^{rot} _{n\mu},b ^{rot} _\mu ;b^{(r)rot} _\mu) ,
\label{rotationalinv}
\eeq
where $b ^{rot} _\mu$ is given by
%(\ref{thetaparity}).
\beq
b_\mu ^{rot} =
\left\{
\begin{array}{ll}
b_{\beta}   & \mbox{for}~~~\mu = \alpha  \\
- b_{\alpha}   & \mbox{for}~~~\mu = \beta  \\
b_{\mu}  & \mbox{otherwise}, \\
\end{array}
\right.
\label{thetarotation}
\eeq
and $b_\mu^{(r)rot}$ is defined similarly.

These behaviors are
the ones we expect in the continuum
if we consider $b_\mu ^{(r)}$ as a parameter representing
an external source.
If we regard $b_\mu ^{(r)}$ as a regularization parameter,
which should be fixed as a function of $b_\mu$,
the allowed choices for $b_\mu ^{(r)}$ are
(A) $b _\mu ^{(r)}=b_\mu$,
(B) $b _\mu ^{(r)} \equiv 0$, and
(C) $b _\mu ^{(r)} \equiv \pi$.
For the 2D U(1) case,
(A) and (B) correspond to the results given by
eqs. (\ref{exactgeneral_m}) and (\ref{exactgeneral_abc}),
respectively.
(C) has not been encountered in Section \ref{Review},
since we started from the known result for $b_\mu =0$,
for which the phase choice corresponding to (C) becomes ill-defined.
As we have seen in Section \ref{BC},
(A) can be singled out by imposing the translational invariance,
but only by sacrificing the property
that a singularity in the gauge configuration can be
absorbed by a change of the boundary condition for the fermion.

%The corresponding relations in the continuum are
%exactly what we expect for any chiral gauge theory.
%The explicit exact result for 2D U(1) chiral gauge theory
%indeed obeys the symmetries.
%Thus, we have found that, if we take $\theta^{(r)}=0$,
%the space-time symmetries of the fermion determinant 
%are preserved manifestly on the lattice.

%\beqa
%\mbox{(i)}&~~~~~&
%b _\mu ^{(r)}=b_\mu \n
%\mbox{(ii)}&~~~~~&
%b _\mu ^{(r)}= 0 \n
%\mbox{(iii)}&~~~~~&
%b _\mu ^{(r)}= \pi \n
%\eeqa
 
\vspace*{1cm}

\section{Singular gauge background}
\label{Singular}
\setcounter{equation}{0}

In Ref. \cite{Anomaly}, it was discovered that 2D U(1) chiral
gauge theories have an anomaly under singular gauge transformations
in general, even if the gauge anomaly for non-singular
gauge transformations is cancelled.
%We argue in the rest of the paper that their conclusion must be
%changed if one takes the breaking of the translational invariance 
%into account.
We first point out that there is actually an ambiguity
in the calculation of chiral determinants in the continuum when
the gauge configuration has a delta-function like singularity.
A typical configuration we consider here is given by
\beq
A^s_\mu (x) = A_\mu (x) + 2 \pi c _\mu \delta (x_\mu - \tilde{x}_\mu) ,
\label{singcont}
\eeq
where $A_\mu (x)$ is a non-singular function and
$c_\mu$ is a real coefficient.
%The ambiguity occurs when we put the configuration
%on the lattice.

Let us consider first regularizing the 
delta-function like singularity as
\beq
A^{reg}_\mu (x) = A_\mu (x) + 2 \pi c _\mu f (x_\mu -\tilde{x}_\mu) ,
\label{regularizedcont}
\eeq
where $f(t)$ is a non-singular function which we send to $\delta (t)$
in the end.
We use (\ref{exactgeneral_r})
for the non-singular gauge configuration $A^{reg}_\mu (x)$ and 
finally take the limit $f(t) \rightarrow \delta (t)$.
The result we obtain in this way is
\beqa
D_{\Ione}(A^s_\mu,b_\mu ; b_\mu ^{(r)})
&=& \ee ^{i \xi} D \left(A_\mu+ \frac{2 \pi c_\mu}{\ell}, b_\mu  
; b _\mu ^{(r)} \right) ,
\label{newresult}
\\
\xi &=& - \frac{1}{2}
\left[ c_1  \left. \int \dd x_2 \del _1 \phi \right|
_{x_1 = \tilde{x}_1}
+ c_2 \left. \int \dd x_1 \del _2  \phi \right| 
_{x_2 = \tilde{x}_2} \right] .
\eeqa
Note that 
%there is no projection function $P$ acting 
%on $c_\mu$ and 
the result is not invariant under $c_\mu \rightarrow 
c_\mu + n_\mu$.

On the other hand, we can treat the singularity in the gauge configuration
just as we did in Section \ref{Review} when
we considered a change of the boundary condition for the fermion
as a singularity in the gauge configuration.

When the singularity resides {\itshape exactly on the boundary},
namely for $\tilde{x}_\mu = 0$, 
the singularity should not be distinguished from the
boundary condition, and therefore we have
\beq
D_{\IItwoa} (A^s _\mu,b_\mu ;b_\mu ^{(r)}) =
D(A_\mu,b_\mu + c_\mu ;b_\mu ^{(r)})  .
\label{eq:absorb}
\eeq
This is the case considered in  Ref. \cite{Anomaly}.
For $\tilde{x}_\mu \neq 0$, one can follow the same steps
as we did in deriving (\ref{exactgeneral_abc}) and arrive at
\beqa
D_{\IItwob}(A^s_\mu,b_\mu;b_\mu ^{(r)}) 
&=& \ee ^{i \zeta} D \left(A_\mu+  \frac{ 2 \pi P[ c_\mu]}{\ell}, 
b_\mu 
;b_\mu ^{(r)} \right)  ,
\label{newresult2}
\\
\zeta &=& - \frac{1}{2}
\left[  P[c_1]  \left. \int \dd x_2 \del _1 \phi 
\right|_{x_1 = \tilde{x}_1}
+ P[c _2]  \left. \int \dd x_1 \del _2  \phi \right| 
_{x_2 = \tilde{x} _2} \right] .
\eeqa
The only difference between (\ref{newresult2})
and (\ref{newresult}) 
is that we now have the projection function $P$
acting on $c_\mu$.
They coincides for $|c _\mu| < 1/2$,
but not in general.
Using the identity (\ref{iden}),
one finds that the difference can occur only in the phase.
Both (\ref{eq:absorb}) and (\ref{newresult2})
are invariant under $c_\mu \rightarrow
c _\mu + n_\mu$ as they should.
(\ref{newresult2}) in the limit of 
$\tilde{x}_\mu  \rightarrow 0$ 
is not necessarily equal to (\ref{eq:absorb}).
One can show that they are equal for the following two cases : 

(i) $b _\mu ^{(r)}= b_\mu $

(ii) $b_\mu ^{(r)} = 0$ and $|P [b _\mu]
+ P [c _\mu ]| < \frac{1}{2} $,

\noindent but not in general.
(i) is expected since the translational invariance
is manifestly preserved
for this choice of $b _\mu ^{(r)}$.
One finds that the discontinuity can occur only in the phase,
again, using the identity (\ref{iden}).
On the other hand, $D_{\Ione}$ given by
(\ref{newresult}) has no discontinuity at $\tilde{x}_\mu =0$ at all.

Thus the continuum result for $D(A_\mu ^s, b _\mu
;b_\mu ^{(r)})$ can be given by
$D_{\Ione}$ or $D_{\IItwo}$
depending on how one treats the singular 
gauge configuration (\ref{singcont}).
$D_{\Ione}$ is the one given by a limit
of (\ref{exactgeneral_r}), but 
$D_{\IItwo}$ cannot be obtained this way.
%One should keep this in mind when one discusses any general
%properties of the theory by simply referring to (\ref{exactgeneral_r}).
We should also note that the ambiguity for singular gauge configurations
is exactly due to the lack of gauge invariance 
for a single Weyl fermion.
As a consequence, the ambiguity lies only in the phase factor.
In vector-like gauge theories,
the singularity, no matter how one puts it on the lattice,
can always be spread out over the whole space-time by
a gauge transformation as we will see in the next section.
Thus the ambiguity for singular gauge configurations as well as 
the translational anomaly is peculiar to 
chiral gauge theories. 
In what follows, we will see that this ambiguity can be 
reproduced by the overlap formalism.

%In the following, we show that both
%$D_{\Ione}$ and $D_{\IItwo}$

%%%%%%%%%%%%%

Within the overlap formalism,
the ambiguity arises when one puts
the continuum singular configurations such as (\ref{singcont})
on the lattice.
$D_{\Ione}$ can be reproduced
by putting $A_\mu ^{reg}$ on the lattice as in (\ref{eq:linkvar}).
After taking the continuum limit, one takes $f(t)$ to $\delta (t)$.
(\ref{newresult}) can be trivially reproduced once
one admits that (\ref{eq:statement_r}) holds
for $U_{n\mu}$ related to a non-singular $A_\mu(x)$
through (\ref{eq:linkvar}).

$D_{\IItwo}$ can be reproduced by
putting the singular gauge configuration (\ref{singcont})
on the lattice as 
\beq
U^s _{n\mu} = 
\left\{
\begin{array}{ll}
\ee^{2 \pi i c _\mu} U_{n\mu}  & \mbox{for}~~~n_\mu = \tilde{n}_\mu  \\
U_{n\mu}  & \mbox{otherwise},
\end{array}
\right.
\label{eq:singularlat}
\eeq
where $U_{n\mu}$ is related to
$A_\mu(x)$ in (\ref{singcont})
through (\ref{eq:linkvar}) and
$\tilde{x}_\mu = \tilde{n}_\mu a$.

\begin{figure}
\begin{center}
  \leavevmode
  \epsfxsize=8cm
  \epsfbox{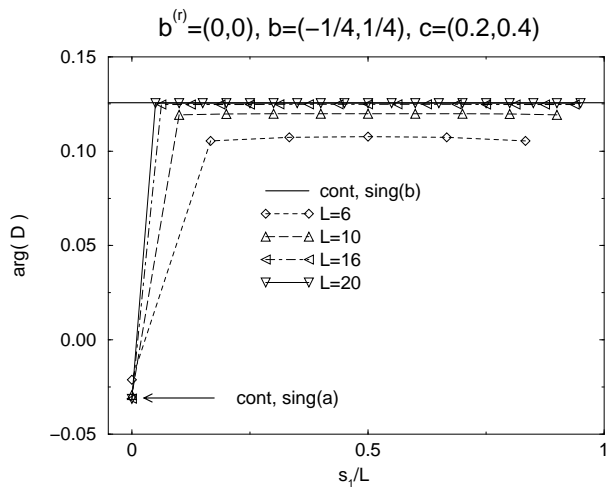}
\end{center}
\caption{
The same plot as in Fig. \ref{fig:TransOsc2} except that
we take the original configuration to be a 
singular gauge configuration,
$U_{n\mu}^s$ given by (\ref{eq:singularlat})
with $U_{n\mu}=1$, $\tilde{n}_\mu = L-1$ 
and $c _\mu = (0.2,0.4)$.
The boundary condition is taken to be $b_\mu = (-1/4,1/4)$.
The boundary condition for the reference state is taken to be
antiperiodic ($b^{(r)}_\mu=0$).
We take the shift to be symmetric in the two directions
($s_1=s_2$), and plot the result
against the shift $s_1/L$ on physical scale.
The arrow with ``sing(a)'' represents the continuum prediction
(\ref{eq:absorb}) for the case in which the singularity resides
exactly on the boundary.
The horizontal line denoted as ``sing(b)'' represents 
the continuum prediction (\ref{newresult2})
for the case in which the singularity resides off the 
boundary.
}
\label{fig:TransSing1}
\end{figure}

There are actually many other ways to put the singular 
gauge configuration on the lattice, between these two extremes,
since one can spread the singularity on two links or
as many links as one likes.
If one naively applies the formula (\ref{eq:linkvar}),
one obtains (\ref{eq:singularlat}).
Note, however, that all the other lattice configurations go to
(\ref{singcont}) in the continuum limit
and therefore can be considered as equally good lattice regularizations
of (\ref{singcont}).
The important point is that the result depends
on which way we adopt for discretizing the singularity (\ref{singcont}).
We also recall that in Section \ref{Review}
we considered a change of the boundary condition for the fermion
as a singularity in the gauge configuration.
There we didn't have this ambiguity
and we put the singularity on a single link,
since boundary conditions should be imposed literally 
{\itshape on the boundary}.
But this is not the case when we discuss singularities
in gauge configurations.

Let us check that $D_{\IItwo}$ given by
(\ref{eq:absorb}) for $\tilde{x}_\mu = 0$ and 
by (\ref{newresult2}) for $\tilde{x}_\mu \neq 0$
can be reproduced by the 
overlap formalism\footnote{In Ref. \cite{Anomaly},
it has been checked numerically that
(\ref{eq:absorb}) can be
reproduced by the overlap formalism for $A_\mu (x)= 0$ and 
$b_\mu = b_\mu ^{(r)} =0$.}.
In Fig. \ref{fig:TransSing1}
we make a plot similar to the one we made 
in Fig. \ref{fig:TransOsc2} 
except that we consider a singular gauge configuration.
We take the original configuration to be (\ref{eq:singularlat}) 
with $U_{n\mu}=1$, $\tilde{n}_\mu = L-1$ and $c _\mu= (0.2,0.4)$.
We take $b_\mu =(-1/4,1/4)$ and $b_\mu ^{(r)} = 0$.
We find that the result expected in the continuum is reproduced 
clearly.
Note again that for $b_\mu ^{(r)} = 0$
the gap at $s_1/L =0$ disappears
when $|P [b _\mu] + P [c_\mu ]| < 1/2$.
If we took the boundary condition to be $b_\mu = (1/4,-1/4)$
with the same $c_\mu$,
the gap would disappear in the continuum limit.
We have also checked this numerically.

\vspace*{1cm}

\section{Constructing anomaly free gauge theories}
\label{anomalyfree}
\setcounter{equation}{0}

In this section, we reconsider the gauge anomaly
under singular gauge transformations discovered in
Ref. \cite{Anomaly}.
%we discuss in what respects the observation
%made in the previous section could be important.
%Here we consider how to construct 
%anomaly free 2D U(1) chiral gauge theories.
%The observations made in the previous sections change
%their conclusion.

There are two kinds of gauge transformation
in the present model.
One is
\beq
A_\mu \rightarrow A_\mu + \del _\mu \Lambda  ,
\label{smallgaugetr}
\eeq
which is a gauge transformation that can be obtained
by repeated use of infinitesimal gauge transformations.
Let us call it a ``small'' gauge transformation.
The other one is 
\beq
A_\mu \rightarrow A_\mu + \frac{2 \pi n_\mu}{\ell} ,
\label{largegaugetr}
\eeq
where $n_\mu$ is an integer vector.
This is what we call a ``large'' gauge transformation,
which has a nontrivial topology and cannot be obtained by
repeated use of infinitesimal gauge transformations.
If we decompose the gauge background $A_\mu$
as in (\ref{Adecompose}),
(\ref{smallgaugetr}) gives $\chi \rightarrow \chi + \Lambda $,
whereas (\ref{largegaugetr}) gives
$h_\mu \rightarrow h_\mu + n_\mu $,
where $h_\mu = \frac{\ell \alpha_\mu}{2 \pi}$ as before.

Let us first restrict ourselves to non-singular gauge configurations.
Then we only have to consider (\ref{exactgeneral_r}).
The exact result (\ref{exactgeneral_r}) for a single Weyl fermion
is not invariant under the above two kinds of gauge transformation,
which is nothing but the gauge anomaly.
In order to cancel the gauge anomaly,
we have to add extra Weyl fermions.
Note that the fermion determinant for a right-handed fermion
is given by (\ref{exactgeneral_r}), and by its complex conjugate
for a left-handed fermion, both with a unit charge.
For charge $q$ fermion, $h_\mu$, $\phi$ and $\chi$ should
be multiplied by $q$ in the corresponding formulae for fermions
with a unit charge.
Let us consider a model with $N_R$ right-handed fermion
with charge $q^R_i$ ($i=1,\cdots,N_R$)
with boundary conditions parametrized by $b^R _{\mu i}$,
and $N_L$ left-handed fermion with charge 
$q^L_i$ ($i=1,\cdots,N_L$)
with boundary conditions parametrized by $b^L _{\mu i}$.
Let us denote the fermion determinant of the whole system as
$D(A_\mu (x) , b ^R_{\mu i},b ^L_{\mu i};
b ^{R(r)}_{\mu i},b ^{L(r)}_{\mu i} )$,
which is nothing but the product of the corresponding fermion
determinant for each fermion.
By referring to the exact results,
one can easily find the condition for
the invariance of
$D(A_\mu (x) , b ^R_{\mu i},b ^L_{\mu i} ;
b ^{R(r)}_{\mu i},b ^{L(r)}_{\mu i} )$
under (\ref{smallgaugetr}) and (\ref{largegaugetr}).
The invariance under a small gauge transformation 
(\ref{smallgaugetr}) requires
\beq
\sum_{i=1}^{N_R} (q^R_i)^2  = \sum_{i=1} ^{N_L} (q^L_i)^2  .
\label{pertanom}
\eeq
The invariance under a large gauge transformation 
(\ref{largegaugetr}) further requires
\beq
\sum _{i=1}^{N_R} q^R_i 
\left(  P [b^R_{i \mu}] -  
 \frac{1}{2}  P [b^R_{i \mu} - b^{R(r)}_{i \mu} ]  \right)
- \sum _{i=1}^{N_L} q^L_i   
\left(  P [b^L_{i \mu}] -  
 \frac{1}{2}  P [b^L_{i \mu} - b^{L(r)}_{i \mu} ]   \right)
=  m_\mu,
\label{largeanom}
\eeq
where $m_\mu$ is an integer vector.
We have used the identity (\ref{iden}).

The condition for translational invariance is given by
\beq
\sum _{i=1}^{N_R} q^R_i 
P [b^R_{i \mu} - b^{R(r)}_{i \mu}] 
- \sum _{i=1}^{N_L} 
q^L_i 
P [b^L_{i \mu} - b^{L(r)}_{i \mu} ]  = 0 .
\label{transinv_cond}
\eeq
This means that
the translational invariance can have an anomaly even if
both (\ref{pertanom}) and (\ref{largeanom}) are satisfied.

Now let us also consider singular gauge configurations
$A^s_\mu$ given by (\ref{singcont}).
In this case, it is not sufficient to consider only 
(\ref{exactgeneral_r}) as we explained in the previous
section.
Therefore, the gauge invariance might be violated even if
(\ref{pertanom}) and (\ref{largeanom}) are satisfied.
Likewise, the translational invariance might be violated even if
(\ref{transinv_cond}) is satisfied.
Indeed, this is what happens.

As stated in Ref. \cite{Anomaly},
$A^s_\mu$ can be gauge-transformed to a 
configuration $A^u_\mu$, which does not
have a delta-function like singularity as
\beqa
A^s_\mu (x) &=& A^u_\mu (x) + \del _\mu \Lambda(x;\tilde{x},c) ,
\label{singgaugetr}
\\
A^u_\mu (x) &\defeq& A_\mu (x) + \frac{2 \pi c_\mu}{\ell} .
\label{uniformgauge}
\eeqa
The transformation function $\Lambda (x;\tilde{x},c)$ is given by
\beq
\Lambda (x;\tilde{x},c) = 
- \frac{2 \pi c_1}{\ell} \Xi (x_1 ;\tilde{x}_1) 
- \frac{2 \pi c_2}{\ell} \Xi (x_2 ;\tilde{x}_2)  ,
\label{transfn}
\eeq
where
\beq
\Xi (t;\tilde{t}) =
\left\{
\begin{array}{ll}
t  & \mbox{for}~~~0<t < \tilde{t} \\
t - \ell & \mbox{for}~~~\tilde{t} < t < \ell .
\end{array}
\right.
\eeq
This gauge transformation is singular in the sense that
$\Lambda (x;\tilde{x},c)$ 
has a discontinuity on $\{ x_1 =\tilde{x}_1 \} 
\cup \{ x_2 =\tilde{x}_2 \}$
as a function on the 2-dimensional torus.
Note also that changing the location of the singularity $\tilde{x}_\mu$ 
can be achieved by successive singular gauge transformations.

In order to construct an anomaly-free chiral gauge theory,
the fermion determinant should be invariant under
such singular gauge transformations as well, namely,
\beq
D(A^s_\mu (x) , b ^R_{\mu i},b ^L_{\mu i} ;
b ^{R(r)}_{\mu i},b ^{L(r)}_{\mu i} )
=
D(A^u_\mu (x) , b ^R_{\mu i},b ^L_{\mu i} ;
b ^{R(r)}_{\mu i},b ^{L(r)}_{\mu i} ) .
\label{singinv}
\eeq
Obviously, the consequence of this requirement
depends on which of $D_{\Ione}$ and $D_{\IItwo}$
we consider as the continuum result for the
singular gauge configuration $A_\mu ^s$.
If we consider $D_{\Ione}$,
(\ref{singinv}) is satisfied automatically,
so long as (\ref{pertanom}) is satisfied,
since $D_{\Ione}$ can be obtained as a limit of (\ref{exactgeneral_r}).

If we consider $D_{\IItwo}$, on the other hand,
the absence of the gauge anomaly under singular gauge transformations
does not automatically follow
from (\ref{pertanom}) and (\ref{largeanom}) and puts
an additional condition which should be satisfied by 
the theory in order to make it anomaly free.
The phase choice of the fermion determinant adopted
in Ref. \cite{Anomaly}
corresponds to taking an antiperiodic boundary condition
for all the fermion species irrespective of the boundary conditions
for these fermions; namely
$b ^{R(r)}_{\mu i} = b ^{L(r)}_{\mu i} = 0$.
The argument was restricted
to the case when the singularity in the gauge configuration resides 
exactly on the boundary.
Referring to (\ref{eq:absorb}), we find that
the condition is
\beqa
&~& D \left(A _\mu (x) + \frac{2 \pi c_\mu}{\ell}, b^R_{\mu i},
b^L_{\mu i} ;
b ^{R(r)}_{\mu i} = 0 ,b ^{L(r)}_{\mu i} = 0 \right)  \n
&=& D(A_\mu (x) , b^R_{\mu i} + q^R_i c_\mu ,
b^L_{\mu i} + q^L_i c_\mu  ;
b ^{R(r)}_{\mu i} = 0 ,b ^{L(r)}_{\mu i} = 0  ) .
\label{singanomaly}
\eeqa
As claimed in Ref. \cite{Anomaly}, this is not always
satisfied even if (\ref{pertanom}) and (\ref{largeanom}) hold.
For example, 
a model with four right-handed fermions with a unit charge
and one left-handed fermion with charge two, 
all of which obeying an antiperiodic
boundary condition, satisfies (\ref{pertanom}) and (\ref{largeanom}),
but not (\ref{singanomaly}).
This led the authors of Ref. \cite{Anomaly} to twist the
boundary conditions as
\beq
b^R_{1 \mu } = \left(\frac{1}{4},\frac{1}{4}\right) ;~~~
b^R_{2 \mu } = \left(\frac{1}{4},-\frac{1}{4}\right) ;~~~
b^R_{3 \mu } = \left(-\frac{1}{4},\frac{1}{4}\right) ;~~~
b^R_{4 \mu } = \left(-\frac{1}{4},-\frac{1}{4}\right) ;~~~
b^L_{1 \mu } = (0,0),
\label{twistedbc}
\eeq
with which one can satisfy (\ref{singanomaly}) as well.
Let us call the former model ``antiperiodic 11112 model''
and the latter ``twisted 11112 model''.

\begin{figure}
\begin{center}
  \leavevmode
  \epsfxsize=8cm
  \epsfbox{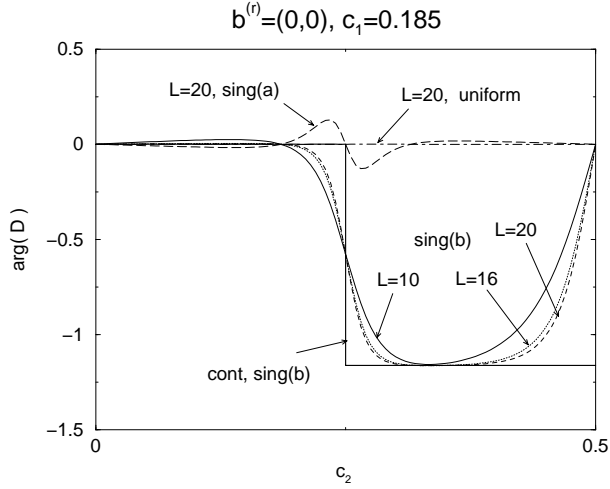}
\end{center}
\caption{
The argument of the normalized overlap determinant 
for the twisted 11112 model.
As singular gauge configurations we take,
$U_{n\mu} ^s$ given by (\ref{eq:singularlat})
with $U_{n\mu}=1$, $c_1=0.185$.
``sing(a)'' represents the results for $\tilde n_\mu=L-1$,
where as 
``sing(b)'' represents the results for $\tilde n_\mu=L/2-1$.
``uniform'' represents the results for the uniform gauge configurations
$U_{n\mu}^u=\ee ^{2 \pi i c_\mu / L}$, which can be obtained by 
a gauge transformation from $U_{n\mu} ^s$.
The horizontal axis represents $c_2$. 
}
\label{fig:Sing.ICG.tot}
\end{figure}

Since the translational invariance is not preserved with
that phase choice of the fermion determinant, however,
one should also consider
the case when the singularity does not coincide with the boundary.
Referring to (\ref{newresult2}),
one finds that (\ref{singinv}) is satisfied if
$| q_i ^L c _\mu| < 1/2 $ and $| q_i ^R c_\mu| < 1/2$
for all $i$, but not in general, even if
(\ref{pertanom}), (\ref{largeanom}) and (\ref{singanomaly})
are satisfied.
This is illustrated in Fig. \ref{fig:Sing.ICG.tot}
for the twisted 11112 model.
As singular gauge configurations, we consider
$U_{n\mu} ^s$ given by (\ref{eq:singularlat})
with $U_{n\mu}=1$, $c_1=0.185$.
``sing(a)'' represents the results for $\tilde n_\mu=L-1$,
whereas 
``sing(b)'' represents the results for $\tilde n_\mu=L/2-1$.
``uniform'' represents the results for the uniform gauge configurations
$U_{n\mu}^u=\ee ^{2 \pi i c_\mu / L}$, which can be obtained by 
a gauge transformation from $U_{n\mu} ^s$.
We plot the argument of the normalized overlap determinant
for the twisted 11112 model as a function of $c_2$.
We only show the results for positive $c_2$.
Note that the fermion determinants for $-c_2$ 
for the three types of configurations we consider 
in Fig. \ref{fig:Sing.ICG.tot}
can be obtained by taking the complex conjugate of those for $c_2$.
This can be shown by 
using the parity transformation property (\ref{parityinv}).
The normalized fermion determinants
for the uniform gauge configurations are
real positive in the continuum as noted in Ref. \cite{Anomaly}.
This is seen to be reproduced by the overlap formalism
in Fig. \ref{fig:Sing.ICG.tot}.
When the singularity coincides with the boundary 
(sing(a) in Fig. \ref{fig:Sing.ICG.tot}), the results
agree with those for the uniform gauge configurations.
A slight discrepancy seen near $c_2=1/4$
can be understood if we look at the fermion determinant for
each species separately.
One finds that the results for the charge-2 left-handed fermion 
and those for the charge-1 right-handed fermions with
$b^R_{i2} = 1/4$ have gaps at $c_2=1/4$
and the gaps are expected to cancel each other in the continuum limit.
Therefore, the discrepancy is considered to be a finite
lattice spacing effect.
On the other hand, 
when the singularity is off the boundary
(sing(b) in Fig. \ref{fig:Sing.ICG.tot}), the results
agree with those for the uniform gauge configurations
for $c_2 < 1/4 $, but not for 
$c_2 > 1/4 $ as expected.
The gap seen at $c_2 = 1/4$ is due to the charge-2 left-handed
fermion, and
the one seen at $c_2 = 1/2$ is due to 
the charge-1 right-handed fermions.
The approach to the continuum prediction is 
slow near these gaps.

\begin{figure}
\begin{center}
  \leavevmode
  \epsfxsize=8cm
  \epsfbox{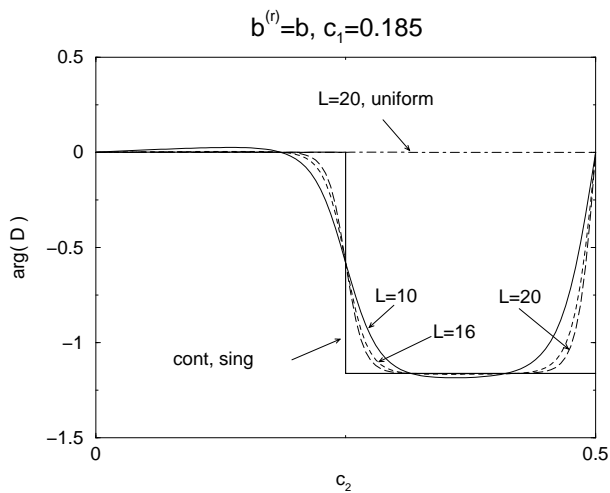}
\end{center}
\caption{
The same plot as in Fig. \ref{fig:Sing.ICG.tot}, except that we take
the boundary condition for the reference state
to be $b_\mu ^{(r)}=b_\mu$.
Since the translational invariance is manifestly preserved on the lattice
for this choice of $b_\mu ^{(r)}$,
the results for the singular gauge configurations,
which are denoted as ``sing'',
do not depend on $\tilde n_\mu$, the location of the singularity.
}
\label{fig:Sing.maint.tot}
\end{figure}

The above argument shows 
that if we consider $D_{\IItwo}$ as the continuum result
for the singular gauge configurations,
even the twisted 11112 model is not completely anomaly-free.
Note also that for the boundary condition (\ref{twistedbc}),
(\ref{transinv_cond}) is satisfied.
However, the translational invariance is broken
as is seen above, because we are considering singular
gauge configurations, which are not considered in
deriving (\ref{transinv_cond}).

When we consider the case in which 
the singularity of the gauge configuration is off
the boundary, the choice of the boundary conditions
makes no difference to the anomaly
for singular gauge transformations.
Actually there is no way to cancel the anomaly 
under singular gauge transformations
except by putting a copy of fermions with the same charge 
and the same boundary condition
but with the opposite chirality, which makes the theory vector-like.
This conclusion does not depend
on the choice of $b _\mu ^{(r)}$.
%A change of $b _\mu ^{(r)}$ results in adding a phase
%proportional to the Polyakov loops, which does not change under
%the singular gauge transformations.
In particular, the conclusion remains unchanged
even if we take the translationally invariant choice 
$b _\mu ^{(r)} = b_\mu$.
%,which is the most natural one.
This is illustrated in Fig. \ref{fig:Sing.maint.tot}.
Due to the translational invariance, the results for
the singular gauge configurations
do not depend
on the location of the singularity.
They agree with the results for the uniform gauge configurations
for $c_2 < 1/4 $, but not for 
$c_2 > 1/4 $.

\vspace*{1cm}

%%%

\section{Generalization to other dimensions}
\label{generalization}
\setcounter{equation}{0}

In this section, we generalize 
our main results to any even dimension for the abelian case.
We discuss the translation anomaly, the ambiguity in the continuum
calculation of chiral determinants for
singular gauge configurations,
and the gauge anomaly under singular gauge transformations.
Although we do not know the exact results for
chiral determinants except in two dimensions,
we can address the above issues
only by using the knowledge of gauge anomaly.

We denote the translationally invariant chiral determinant 
for a single right-handed Weyl fermion 
on a torus of general even dimension as $D(A_\mu , b_\mu)$. 
We define the current $j_\mu [A]$ by
\beq
j_\mu [A] = \frac{\delta}{\delta A_\mu (x)}
\ln D(A_\nu ,b_\nu)  .
\eeq
The anomaly equation is given by
\beq
\del  _\mu j_\mu =  i \del _\mu  K_\mu  [A]  ,
\label{anomalyeq}
\eeq
where $K_\mu [A]$ is defined for $D=2 p$ by
%given for $D=2$ and $D=4$, for example, by
\beq
K_\mu [A] =
\frac{1}{2 (2 \pi)^p p ! } 
\epsilon_{\mu \nu_1 \mu_2 \nu_2 \cdots \mu_p \nu_p}
A_{\nu_1} \del_{\mu_2} A_{\nu_2} \cdots \del _{\mu _p} A_{\nu _p} .
%\left\{
%\begin{array}{ll}
%\frac{1}{4 \pi}  \epsilon _{\mu\nu} A_\nu
%  & \mbox{for}~~~D=2 
%\label{K2D}
%\\
%\frac{1}{8 \pi ^2}  
%\epsilon _{\mu\nu\lambda\rho} A_\nu \del_\lambda A_\rho
%  & \mbox{for}~~~D=4 .
%\end{array}
%\right.
\label{Kmu}
\eeq

Let us consider a singular gauge configuration
$A_\mu ^s (x) $ given by (\ref{singcont}),
where $\tilde{x}_\mu$ and $c_\mu$ 
represent the location and the strength of the singularity,
respectively.
As we discussed in Section \ref{Singular},
the continuum calculation of $D(A_\mu  ^s (x) ,b_\mu)$
has an ambiguity corresponding to 
(\ref{newresult}) and (\ref{newresult2}) in $D=2$.
We denote the two possible results as
$D_{\Ione} (A_\mu  ^s (x) ,b_\mu)$ and $D_{\IItwo}(A_\mu  ^s (x) ,b_\mu)$ 
in the present case.
%Note that Since we are considering the translationally invariant phase
%choice for the chiral determinant here,
%$D_{\IItwo (a)} (A_\mu  ^s (x) ,b_\mu)$ and
%$D_{\IItwo (b)} (A_\mu  ^s (x) ,b_\mu)$ are equal.
Let us first consider
$D_{\Ione} (A_\mu  ^s (x) ,b_\mu)$.
We recall that $A^s_\mu$ can be gauge-transformed to a 
configuration $A^u_\mu$ given by (\ref{uniformgauge}),
which does not have any singularity,
by a gauge transformation (\ref{singgaugetr})
with the transformation function $\Lambda (x;\tilde{x},c)$
given by (\ref{transfn}).
We, therefore, obtain
\beq
\ln D_{\Ione} (A_\mu  ^s (x),b_\mu) - \ln D(A_\mu  ^u (x),b_\mu)
=    
- i  \int \dd ^{D} x ~
\Lambda (x ; \tilde{x}  , c )
 \del _ \mu   K_\mu [A]     ,
%&=&
%- i \int \dd ^{D} x
%\del _ \mu  K_\mu [A]  
%\Lambda (x_\mu ; \tilde{x} _\mu , c_\mu)  ,
\label{lnD}
\eeq
where we have used the fact that 
$ \del  _\mu K_\mu $ is invariant under a gauge transformation 
$A _\mu \rightarrow A_\mu + \del_\mu \Lambda$ and
a constant shift
$A _\mu \rightarrow A_\mu + \alpha _\mu $.
Differentiating both sides of (\ref{lnD})
with respect to $\tilde{x}_\mu$, we obtain
\beq
\frac{\del}{\del \tilde{x}_\mu} 
\ln D_{\Ione} (A_\mu  ^s (x),b_\mu)
=   2 \pi i c_\mu    \left.  \frac{\del}{\del \tilde{x}_\mu} 
 \int \dd ^{D-1} x   ~ K_\mu [A] \right| _{x_\mu = \tilde{x}_\mu} .
\label{generalized}
\eeq
The integral on the r.h.s. of
(\ref{generalized})
yields the Chern-Simons action on the 
boundaries in general. 
The corresponding result for $D_{\IItwo}$ can be obtained
by simply replacing $c_\mu$ by $P[c_\mu]$ in (\ref{generalized}).
\beq
\frac{\del}{\del \tilde{x}_\mu} 
\ln D_{\IItwo} (A_\mu  ^s (x),b_\mu)
=   2 \pi i P[c_\mu]    \left.  \frac{\del}{\del \tilde{x}_\mu} 
 \int \dd ^{D-1} x   ~ K_\mu [A] \right| _{x_\mu = \tilde{x}_\mu} .
\label{generalized2}
\eeq
One can see that these results are consistent with
(\ref{newresult}) and (\ref{newresult2}) in $D=2$
by using (\ref{Kmu}).
(\ref{generalized}) and (\ref{generalized2}) are equal
for $|c _\mu| < 1/2 $, but not in general.

Let us then discuss the translational anomaly.
We define $D(A_\mu, b_\mu ; b_\mu ^{(r)})$ 
as in (\ref{contcounterpart}) and (\ref{Adoubleprime})
in any even dimension.
We note that
\beq
D(A_\mu (x+\tilde{x}),b_\mu;b_\mu ^{(r)}) 
= D_{\IItwo}( A_\mu ^s (x),b_\mu)  ,
\label{trans_sing_rel}
\eeq
where $c_\mu$ in $A_\mu ^s (x)$ should be taken to be
$c_\mu = b_\mu - b_\mu ^{(r)}$.
Therefore, non-vanishing of (\ref{generalized2})
immediately implies the existence of translational anomaly.
The translational anomaly vanishes for $b_\mu ^{(r)} = b_\mu $
as it should.
From this result, we can deduce that
the effect of having $b_\mu ^{(r)}$ different from
$b_\mu$ is essentially given by the phase factor
proportional to the Chern-Simons action on the boundaries
up to an irrelevant constant factor;
namely,
\beqa
D(A_\mu (x), b_\mu; b_\mu ^{(r)})
&\sim & \ee ^{i \beta} D(A_\mu, b_\mu), 
\label{generalD}
\\
\beta &=& 2 \pi  \left. 
\sum_{\mu =1}^{D} 
P[b_\mu - b_\mu ^{(r)}]
\int \dd ^{D-1} x   ~ K_\mu [A] \right| _{x_\mu = 0} .
\eeqa
Note that the extra phase factor does not affect the anomaly
equation (\ref{anomalyeq}), since the Chern-Simons action
is invariant under small gauge transformations.
(\ref{generalD}) is consistent with (\ref{exactgeneral_r}) in $D=2$.

Let us finally discuss the gauge anomaly under
singular gauge transformations.
Here we consider the translationally invariant case
$b_\mu ^{(r)} = b_\mu $.
The conclusion is the same for 
$b_\mu ^{(r)} \neq b_\mu $ as long as the singularity
is off the boundary.
We consider the singular gauge configuration $A_\mu ^s (x)$.
We recall that changing the location of the singularity
is a gauge transformation, which can be
achieved by successive singular gauge transformations.
Eqs. (\ref{generalized}) and (\ref{generalized2}), therefore, give
the gauge anomaly for a single right-handed Weyl fermion
under this gauge transformation.
Let us then consider an anomaly-free fermion content which satisfies 
a charge relation corresponding to (\ref{pertanom}) in 2D
and see whether the above gauge anomaly is cancelled as well.
When we consider $D _{\Ione}$, the gauge anomaly
(\ref{generalized}) automatically cancels due to the charge relation.
When we consider $D_{\IItwo}$, however,
the gauge anomaly (\ref{generalized2}) cancels 
when $|q_i ^L c _\mu| < 1/2 $ and $|q_i ^R c_\mu| < 1/2$,
but not in general.

%Thus we have seen that 
%the translational anomaly, the ambiguity
%in the continuum calculation of the fermion determinant for a
%singular gauge configuration, and the singular gauge anomaly
%exist in any even dimension in the abelian case.

Thus we have seen that 
U(1) chiral gauge theories on a torus in any even dimension
have the translational anomaly and the singular gauge anomaly,
and that the latter suffers from an ambiguity in the continuum
calculation of chiral determinants for singular gauge configurations.

\vspace*{1cm}

\section{Summary and Discussions}
\label{Summary}
\setcounter{equation}{0}

In this paper, we pointed out that
the fermion determinant in chiral gauge theories on a two-dimensional 
torus can have a phase ambiguity,
which is proportional to the Polyakov loops along the boundaries.
The continuum results have been reproduced by the overlap formalism,
where the phase ambiguity comes from the ambiguity of the formalism,
which lies in the boundary condition 
for the reference state used in the Wigner-Brillouin phase choice.
The anomaly equation is not affected by the ambiguous phase factor 
and therefore cannot fix the ambiguity.
Space-time symmetries, such as chirality interchange, 
parity transformation, charge conjugation and 90$^\circ$ rotation
almost fix the ambiguity, but not completely.
One can fix it completely by imposing the translational invariance,
which corresponds, in the overlap formalism,
to taking the boundary condition for the reference state 
to be identical to the one for the fermion under consideration.
But then the fermion determinant loses the property
that a singularity in the gauge configuration
can be absorbed by a change of the boundary condition for the fermion.
The conflict between the translational invariance and the 
absorption of the gauge field singularity by a change of 
the boundary condition,
is realized by the overlap formalism
on the lattice for general chiral gauge theories on a torus.
%Thus, an interesting feature of the phase ambiguity is that
%the requirement that a singularity in the gauge configuration
%can be absorbed by a change of the boundary condition for 
%the fermion, is not compatible with translational
%invariance.

Using these new insights, 
we reconsidered the gauge anomaly under singular gauge transformations
discovered in Ref. \cite{Anomaly}.
We pointed out that the phase choice adopted there
actually corresponds to the one which breaks translational invariance
and that the argument was restricted to the case in which
the singularity in the gauge field resides exactly on the boundary.
If one shifts the singularity off the boundary,
the result changes because of the lack of translational invariance.
Taking this case into account, there is no way to 
make completely anomaly-free 2D U(1) chiral gauge theories
other than to make it vector-like.
The conclusion remains the same no matter how one fixes the 
phase ambiguity of the fermion determinant,
and in particular, it remains unaltered for the 
%most natural phase choice which gives a
translationally invariant phase choice.

We have 
generalized our results to any even dimension for the abelian case.
We have shown that the translational anomaly 
and the anomaly under singular gauge transformations
can be calculated in the continuum only 
with the knowledge of gauge anomaly,
and that they can be expressed 
in terms of the Chern-Simons action on the boundaries in general.
It is natural to expect that 
these anomalies can be reproduced
by the overlap formalism in any even dimension.

The existence of the singular gauge anomaly
poses an apparent contradiction to the fact
that there is an explicit construction
of lattice U(1) chiral gauge theory,
which is gauge invariant on the lattice \cite{Luescher}.
We clarified it by pointing out an ambiguity in
the continuum calculation
of the chiral determinant
for gauge configurations with a delta-function like singularity.
Due to the ambiguity,
the continuum calculations alone cannot
tell whether the singular gauge anomaly is a property which
any lattice regularization of chiral gauge theories
should exhibit, although it is a property of the overlap formalism with
the Wigner-Brillouin phase choice.
%We therefore conclude that
%the anomaly under singular gauge transformations
%is a property of the overlap formalism with
%the Wigner-Brillouin phase choice, but not a 
%property required in the continuum.
%Our conclusion suggests that one will face with an obstacle
%when one tries to apply the construction to the 2D case.
It is interesting to see explicitly how the formalism of 
Ref. \cite{Luescher} avoids the singular gauge anomaly.

Whether the singular gauge anomaly 
is a real obstacle of the overlap formalism 
when one performs the path integral
over the gauge field without gauge fixing is a nontrivial
dynamical question.
In Ref. \cite{Anomaly} it was suggested
that the overlap formalism with the
gauge averaging procedure works for the 2D twisted 11112 model
\cite{Composite,Finite,MCeval,proceedings}
but not for the 2D antiperiodic 11112 model.
This might be the case, but the distinction of the two models,
which was claimed to be the absence of the singular gauge anomaly
for the former, is not true, as we have seen.
This issue also needs further studies.
Even if it turns out that the overlap formalism with
averaging over the gauge orbit does not work for 
{\itshape general} anomaly-free chiral gauge theories,
using a gauge fixing with the formalism might be a
promising approach.

Considering that the overlap formalism 
can be applied to general chiral gauge theories
successfully at least for smooth gauge backgrounds,
it would be nice to reformulate it
in terms of a more standard Lagrangian formalism.
The key to this step is to understand
the profound meaning of the Wigner-Brillouin phase choice.
We expect that our finding concerning
the choice of the reference state
might provide a clue to this problem.
%Still, the fact that the overlap formalism reproduce
%the fermion determinant for fixed background gauge configurations
%as we established in (\ref{eq:statement}) 
%for arbitrary boundary conditions in the 2D U(1) case
%is worth noting.
%This means that the gauge anomaly goes away
%in the continuum limit in the sense that the overlap chiral 
%determinants for any gauge equivalent continuum configurations
%$A_\mu$ give the same result.
%We expect that this is the case for general chiral gauge theories.
%If this is the case,
%the overlap formalism with some kind of gauge fixing 
%might well be a candidate for constructing a general anomaly
%free lattice chiral gauge theory.
%This possibility should be pursued further.
We also hope that the peculiar properties of chiral gauge theories 
elucidated in this paper will be useful 
in constructing any sensible regularizations
of chiral gauge theories as well as revealing interesting 
dynamical features of these theories in the near future.
%It is natural to expect that the overlap gives 
%the correct results if $A_\mu (x)$ is
%finite at any point of $x$.
%If this is the case, we can obtain in principle the correct result
%for the gauge field which has a delta-function like singularity in general,
%by treating 
%such a configuration by a limit series of non-singular configurations.

\vspace*{1cm}

\section*{Acknowledgements}
%\vspace*{1cm}
%\noindent
We would like to thank Y. Kikukawa
for helpful communications
and W. Bietenholz for carefully reading the manuscript.
T.I. is grateful to S. Aoki,  K-I. Nagai, Y. Taniguchi and A. Ukawa
for illuminating discussions and patient encouragements.
J.N. is benefitted from discussions with
J. Ambj\o rn, W. Bietenholz, P. Orland and G. Semenoff.
This work is supported in part by
Grant-in-Aid for Scientific Research 
from the Ministry of Education, Science and Culture
(No. 2375).
T.I. is a Research Fellow of Japan Society for the Promotion of Science.
J.N. is a JSPS Postdoctoral Fellow for Research Abroad.
%supported by Japan Society for the Promotion of Science
%as a Research Fellow Abroad.

\end{document}